\documentclass[epjCONF,final]{svjour}

\usepackage{graphicx}
\usepackage[latin1]{inputenc}

\usepackage{savesym}
\usepackage{amsmath}
\savesymbol{iint}
\usepackage[varg]{txfonts}
\restoresymbol{TXF}{iint}

\usepackage{amsfonts}
\usepackage{amssymb}
\usepackage{bm}
\usepackage{dsfont}
\usepackage{units}

\newcommand{\vect}[1]{{\mbox{\boldmath $#1$}}}
\newcommand{\conjg}[1]{\ensuremath{\hspace{1pt}\overline{\hspace{-1pt}#1\hspace{-1pt}}}\hspace{1pt}}

\def\Slash#1{\setbox0=\hbox{$#1$} % set a box for #1
\dimen0=\wd0 % and get its size
\setbox1=\hbox{/} \dimen1=\wd1 % get size of /
\ifdim\dimen0>\dimen1 % #1 is bigger
\rlap{\hbox to \dimen0{\hfil/\hfil}} % so center / in box
#1 % and print #1
\else % / is bigger
\rlap{\hbox to \dimen1{\hfil$#1$\hfil}} % so center #1
/ % and print /
\fi}

\def\longlongrightarrow{
\relbar\joinrel\relbar\joinrel\relbar\joinrel\relbar\joinrel\rightarrow}

\session-title{%
19$^{\textnormal{\footnotesize th}}$ International %
IUPAP Conference on Few-Body Problems in Physics%
}

\begin{document}

\title{ Covariant solution of the three-quark problem in quantum field theory: \\ the nucleon }
\author{ G. Eichmann\inst{1,2}\fnmsep\thanks{\email{gernot.eichmann@physik.tu-darmstadt.de}}
         \and R. Alkofer\inst{2}
         \and A. Krassnigg\inst{2}
         \and D. Nicmorus\inst{2} }

\institute{Institute for Nuclear Physics, Darmstadt University of Technology, 64289 Darmstadt, Germany
           \and Institut f\"ur Physik, Karl-Franzens-Universit\"at Graz, 8010 Graz, Austria }

\abstract{  We provide details on a recent solution of the nucleon's covariant Faddeev equation in an explicit three-quark approach.
            The full Poincar\'e-covariant structure of the three-quark amplitude is implemented through
            an orthogonal basis obtained from a partial-wave decomposition.
            We employ a rainbow-ladder gluon exchange kernel which allows for a comparison with meson
	    Bethe-Salpeter and baryon quark-diquark studies.
        We describe the construction of the three-quark amplitude in full detail and compare it
        to a notation widespread in recent publications.
        Finally, we discuss first numerical results for the nucleon's amplitude. }

\maketitle

%%%%%%%%%%%%%%%%%%%%%%%%%%%%%%%%%%%%%%%%%%%%%%%%%%%%%%%%%%%%%%%%%%%%%%%%%%%%%%%%%%%%%%%%%%%%%%%%%%%%%%%%%%%%%%%%%%%%%%%%%%%%%%%%%%%%%%%%%%%%%%%%%%%%%%%%%%%%%%%%%%%%%%%%%%%%%%%
%%%%%%%%%%%%%%%%%%%%%%%%%%%%%%%%%%%%%%%%%%%%%%%%%%%%%%%%%%%%%%%%%%%%%%%%%%%%%%%%%%%%%%%%%%%%%%%%%%%%%%%%%%%%%%%%%%%%%%%%%%%%%%%%%%%%%%%%%%%%%%%%%%%%%%%%%%%%%%%%%%%%%%%%%%%%%%%
%%%%%%%%%%%%%%%%%%%%%%%%%%%%%%%%%%%%%%%%%%%%%%%%%%%%%%%%%%%%%%%%%%%%%%%%%%%%%%%%%%%%%%%%%%%%%%%%%%%%%%%%%%%%%%%%%%%%%%%%%%%%%%%%%%%%%%%%%%%%%%%%%%%%%%%%%%%%%%%%%%%%%%%%%%%%%%%
%%%%%%%%%%%%%%%%%%%%%%%%%%%%%%%%%%%%%%%%%%%%%%%%%%%%%%%%%%%%%%%%%%%%%%%%%%%%%%%%%%%%%%%%%%%%%%%%%%%%%%%%%%%%%%%%%%%%%%%%%%%%%%%%%%%%%%%%%%%%%%%%%%%%%%%%%%%%%%%%%%%%%%%%%%%%%%%

\section{Introduction} \label{EichmannG_Introduction}

            Abundant information on the structure of the nucleon is available from experiments,
            and here especially from electroweak probes at all energy scales.
            It is presently both an issue of paramount importance and a formidable theoretical
            challenge in contemporary hadron physics to understand
            the nucleon and its structure in terms of quarks and
            gluons which are the elementary degrees of freedom of quantum
            chromodynamics (QCD), i.e. the
            quantum field theory of the strong interaction.

            The formalism which has been developed to treat the three-body bound-state problem has a longstanding history which dates back to the original work by Faddeev \cite{Faddeev:1960su}.
            Non-relativistic Faddeev equations have found widespread application in the description of three-nucleon systems, see Ref.~\cite{Gloeckle:1995jg} for an overview.
            The covariant generalization of the Faddeev equation to the three-body analogue of a two-body Bethe-Salpeter equation (BSE) \cite{Salpeter:1951sz}
            was formulated in Refs. \cite{Taylor:1966zza,Boehm:1976ya}; a comprehensive introduction can be found in \cite{Loring:2001kv}.
            Within such a framework, the equation describes the baryon as a bound state of three spin-$\nicefrac{1}{2}$ valence quarks
            where the interaction kernel comprises two- and three-quark contributions.

            The formalism of QCD's Dyson-Schwinger equations (for recent reviews, see e.g.~\cite{Fischer:2006ub,Roberts:2007jh})
            provides a way to embed the covariant three-quark Faddeev equation in a consistent quantum-field theoretical setup.
            The dynamical ingredients in the equation can then
            be treated in perfect correspondence with studies of quark and meson properties as well as related aspects of QCD.
            A solution of the equation relies upon knowledge of the dressed quark propagator and the three-quark kernel as well as
            a specification of the Poincar\'e-covariant baryon amplitude.
            The relativistic spin structure of the latter has been explored in \cite{Machida:1974xw,Henriques:1975uh}
            and described in the light-front formalism in \cite{Weber:1986qw,Beyer:1998xy,Karmanov:1998jp,Sun:2001ir}.
            A complete classification according to the Lorentz group and the permutation group $\mathbb{S}_3$ was derived in \cite{Carimalo:1992ia} in terms of covariant three-spinors.

            The biggest obstacle on the way to a direct numerical solution of the three-body bound-state equation is its complexity.
            Upon implementing perturbative quark propagators
            it has been studied, for instance, in the works of Refs.\,\cite{Kielanowski:1979eb,Falkensteiner:1981ab}, in the context of
            a three-body spectator approximation \cite{Stadler:1997iu}, or a Salpeter equation with instantaneous forces \cite{Loring:2001kv}.
            The corresponding equation of a scalar three-particle system with scalar two-body exchange based on the Wick-Cutkosky model
            \cite{Wick:1954eu,Cutkosky:1954ru} was recently investigated and compared to the light-front approach \cite{Karmanov:2008bx}.

            A different kind of simplification can be achieved by considering diquark correlations in the nucleon, see e.g.~\cite{Anselmino:1992vg}
            for an overview. While maintaining full Poincar\'e covariance, the quark-di\-quark model traces the
            nucleon's binding to the formation of colored scalar- and axialvector diquarks, thereby simplifying the Faddeev equation to a quark-diquark
            BSE. This strategy has been applied to investigate nucleon and $\Delta$ properties
            \cite{Hellstern:1997pg,Oettel:2000jj,Alkofer:2004yf,Holl:2005zi,Eichmann:2007nn,Eichmann:2008ef,Nicmorus:2008vb,Eichmann:2008kk,Nicmorus:2008eh}.
	    For more detailed studies of diquark properties in this approach, see \cite{Maris:2002yu,Maris:2004bp}. Further support
	    for the diquark concept has been provided by a study of diquark confinement in Coulomb-gauge QCD \cite{Alkofer:2005ug}.

            In a recent study we have reported the first fully Poincar\'e-covariant computation of the nucleon's mass and Faddeev amplitude beyond
            the quark-diquark approximation \cite{Eichmann:2009qa}. The numerical solution of the Faddeev equation is performed after
            truncating the interaction kernel to a dressed gluon-ladder exchange between any two quarks,
            thereby enabling a direct comparison with corresponding meson studies
            as well as earlier investigations of baryons in the quark-diquark model.
            In the present work we provide details of that calculation with regard to the covariant structure of the nucleon amplitude
            and its decomposition into Dirac tensors, as well as details on the numerical solution of the Faddeev equation.

%           \bigskip\bigskip   % feel free to remove this - or fill the gap by adding more text to the introduction.

            \begin{figure*}[tbp]
                    \begin{center}
                    \includegraphics[scale=0.45]{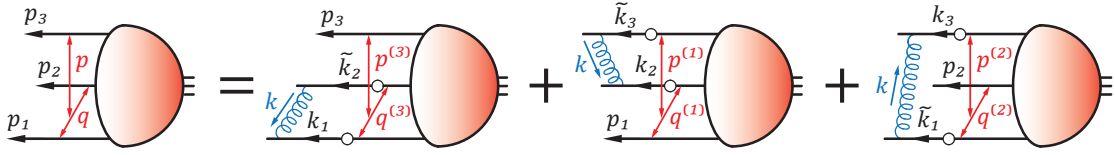}
                    \caption{Faddeev equation \eqref{EichmannG_faddeev:eq} in rainbow-ladder truncation.}\label{EichmannG_fig:faddeev}
                    \end{center}
            \end{figure*}

%%%%%%%%%%%%%%%%%%%%%%%%%%%%%%%%%%%%%%%%%%%%%%%%%%%%%%%%%%%%%%%%%%%%%%%%%%%%%%%%%%%%%%%%%%%%%%%%%%%%%%%%%%%%%%%%%%%%%%%%%%%%%%%%%%%%%%%%%%%%%%%%%%%%%%%%%%%%%%%%%%%%%%%%%%%%%%%
%%%%%%%%%%%%%%%%%%%%%%%%%%%%%%%%%%%%%%%%%%%%%%%%%%%%%%%%%%%%%%%%%%%%%%%%%%%%%%%%%%%%%%%%%%%%%%%%%%%%%%%%%%%%%%%%%%%%%%%%%%%%%%%%%%%%%%%%%%%%%%%%%%%%%%%%%%%%%%%%%%%%%%%%%%%%%%%
%%%%%%%%%%%%%%%%%%%%%%%%%%%%%%%%%%%%%%%%%%%%%%%%%%%%%%%%%%%%%%%%%%%%%%%%%%%%%%%%%%%%%%%%%%%%%%%%%%%%%%%%%%%%%%%%%%%%%%%%%%%%%%%%%%%%%%%%%%%%%%%%%%%%%%%%%%%%%%%%%%%%%%%%%%%%%%%
%%%%%%%%%%%%%%%%%%%%%%%%%%%%%%%%%%%%%%%%%%%%%%%%%%%%%%%%%%%%%%%%%%%%%%%%%%%%%%%%%%%%%%%%%%%%%%%%%%%%%%%%%%%%%%%%%%%%%%%%%%%%%%%%%%%%%%%%%%%%%%%%%%%%%%%%%%%%%%%%%%%%%%%%%%%%%%%

\section{Faddeev amplitude and equation} \label{EichmannG_sec:faddeeveq}

            In QCD baryons appear as poles in the three-quark scattering matrix.
            This allows one to derive
            a relativistic three-body bound-state equation:
            \begin{equation}
                \Psi = \widetilde{K}_\text{(3)}\,\Psi\,, \qquad \widetilde{K}_{(3)} = \widetilde{K}_\text{(3)}^\text{irr} + \sum_{a=1}^3 \widetilde{K}^{(a)}_{(2)}\,,
            \end{equation}
            where $\Psi$ is the bound-state amplitude defined on the baryon mass shell. % $P^2=-M^2$.
            The three-body kernel $\widetilde{K}_{(3)}$
            comprises a three-quark irreducible contribution and the sum of permuted two-quark kernels
            whose quark-antiquark analogues appear in a meson BSE.
            The subscript $a$ denotes the respective accompanying spectator quark.

            The observation of a strong attraction in the $SU(3)_C$ antitriplet $qq$ channel
            has been the guiding idea for the quark-diquark model, namely that quark-quark correlations
            provide important binding structure in baryons.
            This motivates the omission of the three-body irreducible contribution from the full three-quark kernel.
             The resulting covariant Faddeev equation includes a sum of permuted $qq$ kernels (cf.~Fig.\,\ref{EichmannG_fig:faddeev}):
                \begin{equation}\label{EichmannG_faddeev:eq}
                    \Psi_{\alpha\beta\gamma\delta}(p,q,P)  =\sum_{a=1}^3   \int\limits_k    \widetilde{K}_{\alpha\alpha'\beta\beta'\gamma\gamma'}^{(a)} \, \Psi_{\alpha'\beta'\gamma'\delta}(p^{(a)},q^{(a)},P)\,,
                \end{equation}
            where $\widetilde{K}^{(a)}$ denotes the renormalization-group invariant products of a $qq$ kernel and two dressed quark propagators:
            \begin{equation} \label{EichmannG_KSS}
                \widetilde{K}_{\alpha\alpha'\beta\beta'\gamma\gamma'}^{(a)} = \delta_{\alpha\alpha'}\mathcal{K}_{\beta\beta''\gamma\gamma''}\, S_{\beta''\beta'}(k_b)  \, S_{\gamma''\gamma'} (\widetilde{k}_c)\,.
            \end{equation}
            Here, $\{a,b,c\}$ is an even permutation of $\{1,2,3\}$ and linked to the respective Dirac index pairs.

            The spin-momentum part of the full Poincar\'e-covariant nucleon amplitude $\Psi_{\alpha\beta\gamma\delta}(p,q,P)$
            is a spin-$\nicefrac{1}{2}$ four-point function
            with positive parity and positive energy:
            it carries three spinor indices $\{\alpha,\beta,\gamma\}$ for the involved valence quarks
            and one index $\delta$ for the spin-$1/2$ nucleon.
            The amplitude depends on three quark momenta $p_1$, $p_2$, $p_3$ which can be reexpressed in terms of
            the total momentum $P$ and two relative Jacobi momenta $p$ and $q$, where $P^2 = -M^2$ is fixed.
            They are related via (cf. Fig.\,\ref{EichmannG_fig:faddeev}):
                \begin{equation}
                \begin{array}{l}  p = (1-\zeta)\,p_3 - \zeta\,p_d\,,  \\[0.2cm]
                                  q = \dfrac{p_2-p_1}{2}\,,  \\[0.2cm]
                                  P = p_1+p_2+p_3\,,
                \end{array} \quad
                \begin{array}{l}  p_1 =  -q -\dfrac{p}{2} + \dfrac{1-\zeta}{2}\, P\,, \\[0.2cm]
                                  p_2 =  q -\dfrac{p}{2} + \dfrac{1-\zeta}{2}\, P\,, \\[0.2cm]
                                  p_3 =  p + \zeta\, P\,,
                \end{array}
                \end{equation}
            where we abbreviated $p_d:=p_1+p_2$.
            We have chosen equal momentum partitioning $1/2$ for the relative momentum $q$
            and use the value $\zeta=1/3$ in connection with the momentum $p$;
            this value maximizes the upper boundary for the nucleon mass with respect to restrictions arising from the quark propagator's singularity structure \cite{Eichmann:2009zx}.
            The quark propagators $S$ depend on the internal quark momenta
            $k_i=p_i-k$ and $\widetilde{k}_j=p_j+k$.
            The internal relative momenta are given by:
    \renewcommand{\arraystretch}{1.2}
                \begin{equation}
                \begin{array}{l}  p^{(1)} = p+k\,, \\
                                  p^{(2)} = p-k\,, \\
                                  p^{(3)} = p\,,
                \end{array} \qquad\quad
                \begin{array}{l}  q^{(1)} = q-k/2\,, \\
                                  q^{(2)} = q-k/2\,, \\
                                  q^{(3)} = q+k\,.
                \end{array}
                \end{equation}

            The nucleon amplitude can be decomposed into 64 Dirac structures:
                \begin{equation}\label{EichmannG_faddeev:amp}
                    \Psi_{\alpha\beta\gamma\delta}(p,q,P) = \sum_{k=1}^{64} f_k(p^2, q^2, \{z\}) \,\tau_k(p,q,P)_{\alpha\beta\gamma\delta}\,,
                \end{equation}
            where the amplitude dressing functions $f_k$ depend on the five Lorentz-invariant combinations
            \begin{equation} \label{EichmannG_mom-variables}
                p^2\,, \quad q^2\,,\quad   z_0=\widehat{p_T}\cdot\widehat{q_T} \,,\quad z_1 = \hat{p}\cdot\hat{P} \,,\quad  z_2 = \hat{q}\cdot\hat{P}\,.
            \end{equation}
            Here, a hat denotes a normalized 4-vector and $p_T^\mu = T^{\mu\nu}_P p^\nu$ a transverse projection with
	    respect to any four momentum $P$: $T^{\mu\nu}_P = \delta^{\mu\nu} - \hat{P}^\mu \hat{P}^\nu$.
            We abbreviated the angular variables by the shorthand notation $\{z\}=\{z_0,z_1,z_2\}$.
            The Dirac structures $\tau_k(p,q,P)$ will be explained in Section~\ref{EichmannG_sec:nucleonamp}.

%%%%%%%%%%%%%%%%%%%%%%%%%%%%%%%%%%%%%%%%%%%%%%%%%%%%%%%%%%%%%%%%%%%%%%%%%%%%%%%%%%%%%%%%%%%%%%%%%%%%%%%%%%%%%%%%%%%%%%%%%%%%%%%%%%%%%%%%%%%%%%%%%%%%%%%%%%%%%%%%%%%%%%%%%%%%%%%
%%%%%%%%%%%%%%%%%%%%%%%%%%%%%%%%%%%%%%%%%%%%%%%%%%%%%%%%%%%%%%%%%%%%%%%%%%%%%%%%%%%%%%%%%%%%%%%%%%%%%%%%%%%%%%%%%%%%%%%%%%%%%%%%%%%%%%%%%%%%%%%%%%%%%%%%%%%%%%%%%%%%%%%%%%%%%%%
%%%%%%%%%%%%%%%%%%%%%%%%%%%%%%%%%%%%%%%%%%%%%%%%%%%%%%%%%%%%%%%%%%%%%%%%%%%%%%%%%%%%%%%%%%%%%%%%%%%%%%%%%%%%%%%%%%%%%%%%%%%%%%%%%%%%%%%%%%%%%%%%%%%%%%%%%%%%%%%%%%%%%%%%%%%%%%%
%%%%%%%%%%%%%%%%%%%%%%%%%%%%%%%%%%%%%%%%%%%%%%%%%%%%%%%%%%%%%%%%%%%%%%%%%%%%%%%%%%%%%%%%%%%%%%%%%%%%%%%%%%%%%%%%%%%%%%%%%%%%%%%%%%%%%%%%%%%%%%%%%%%%%%%%%%%%%%%%%%%%%%%%%%%%%%%

    \section{Rainbow-ladder truncation} \label{EichmannG_sec:RL}

        To proceed with the numerical solution of the Faddeev equation, we need to specify the quark-quark kernel $\mathcal{K}$
        and the dressed quark propagator $S(p)$ which appear in Eq.\,\eqref{EichmannG_KSS}.
        They are related via the
        axial-vector Ward-Takahashi identity (AXWTI) which encodes the properties of chiral symmetry in connection with QCD.
        Its satisfaction by the interaction kernels in related equations guarantees the correct implementation of
        chiral symmetry and its dynamical breaking, leading e.g.~to a generalized Gell-Mann--Oakes--Renner relation
        valid for all pseudoscalar mesons and all current-quark masses \cite{Maris:1997hd,Holl:2004fr}.
        In particular the pion, being the Goldstone boson related to dynamical chiral symmetry breaking, becomes massless
        in the chiral limit, independent of the details of the interaction. Specifically, we describe the $qq$ kernel
        by a dressed gluon-ladder exchange:
                \begin{equation}\label{EichmannG_RL:kernel}
                    \mathcal{K}_{\alpha\alpha'\beta\beta'}(k) = Z_2^2 \,\frac{4\pi\alpha(k^2)}{k^2} \, T^{\mu\nu}_k \,\gamma^\mu_{\alpha\alpha'} \,\gamma^\nu_{\beta\beta'}\,,
                \end{equation}
            where $k$ is the gluon momentum and $Z_2$ the quark renormalization constant.
            By virtue of the AXWTI, the kernel must also appear in the quark Dyson-Schwinger equation whose solution
            defines the renormalized dressed quark propagator:
            \begin{equation}\label{EichmannG_RL:QuarkDSE}
                S^{-1}_{\alpha\beta}(p) = Z_2 \left( i\Slash{p} + m \right)_{\alpha\beta}  + \int_q \mathcal{K}_{\alpha\alpha'\beta'\beta}(k) \,S_{\alpha'\beta'}(q)\,,
            \end{equation}
            where the bare quark mass $m$ enters as an input and $k=p-q$.
            The inherent color structure of the kernel leads to prefactors $2/3$ and $4/3$ for the integrals in Eqs.\,\eqref{EichmannG_faddeev:eq} and \eqref{EichmannG_RL:QuarkDSE}, respectively.

            Eqs.\,(\ref{EichmannG_RL:kernel}--\ref{EichmannG_RL:QuarkDSE}) define the rainbow-ladder (RL) truncation
            which has been extensively used in DSE studies of mesons and baryons in the quark-diquark model, e.\,g.~\cite{Krassnigg:2009zh,Eichmann:2007nn} and references therein.
           The non-perturbative dressing of the gluon propagator and the quark-gluon
	       vertex are absorbed into an effective coupling $\alpha(k^2)$ for which we adopt the ansatz~\cite{Maris:1999nt,Eichmann:2008ae}
        \begin{equation}\label{EichmannG_couplingMT}
            \alpha(k^2) = \pi \eta^7  \left(\frac{k^2}{\Lambda^2}\right)^2 \!\! e^{-\eta^2 \left(\frac{k^2}{\Lambda^2}\right)} + \alpha_\text{UV}(k^2) \,.
        \end{equation}
        The second term
        \begin{equation}
            \alpha_\text{UV}(k^2) = \frac{ \pi\gamma_m\,\left( 1-e^{-k^2/\Lambda_t^2} \right) } {\ln\sqrt{ e^2 -1 + \left( 1+k^2/\Lambda_{QCD}^2 \right)^2 } }\,,
        \end{equation}
        where $\gamma_m=12/(11 N_C-2 N_f)$ is the anomalous dimension of the quark propagator,
        reproduces the logarithmic decrease of QCD's one-loop perturbative running coupling and vanishes at $k^2=0$.
        In our calculation we use $\Lambda_{QCD}=0.234$ GeV and $\gamma_m=12/25$ which corresponds to $N_f=4$.
        The infrared contribution (the first term in \eqref{EichmannG_couplingMT}) is parametrized by an infrared scale $\Lambda$ and a dimensionless parameter $\eta$
        and yields the non-pertur\-bative enhancement at small and intermediate gluon
	    momenta necessary to generate dynamical chiral symmetry breaking and hence a constituent-quark mass scale.
        ($\{\Lambda,\,\eta\}$ and the infrared parameters used in \cite{Eichmann:2008ae} are related by
        $\mathcal{C}= (\Lambda/\Lambda_t)^3$ and $\omega = \eta^{-1} \Lambda/\Lambda_t$, with $\Lambda_t = 1$ GeV.)

        Beyond the present truncation, corrections arise from pseudoscalar meson-cloud contributions which
        provide a substantial attractive contribution to the `quark core' of dynamically
	    generated hadron observables in the chiral regime and vanish with increasing current-quark mass,
        but also from non-resonant contributions due to the infrared structure of the quark-gluon vertex.
        To anticipate such corrections we exploit the freedom in adjusting the input scale $\Lambda$.
        We adopt two different choices established in the literature in the context of $\pi$ and $\rho$
        properties \cite{Eichmann:2008ae}:

               Setup A is determined by a fixed scale $\Lambda = 0.72$ GeV, chosen in \cite{Maris:1999nt} to reproduce
               the experimental pion decay constant and the phenomenological quark condensate. Corresponding results
               are therefore aimed in principle at a comparison to experimental data for meson and baryon properties
               (see \cite{Krassnigg:2009zh,Nicmorus:2008vb} and references therein).
               Setup B defines a current-mass dependent scale which is deliberately inflated close to the chiral limit,
               where $\Lambda \approx 1$ GeV \cite{Eichmann:2008ae}. It is meant to describe a hadronic quark core which
               must subsequently be dressed by pion-cloud effects and other corrections. As a result, $\pi$, $\rho$,
               $N$ and $\Delta$ observables are consistently overestimated, but (with the exception of the $\Delta$-baryon)
               compatible with quark-core estimates from quark models and chiral perturbation theory (for a detailed
               discussion, see \cite{Eichmann:2008ae,Eichmann:2008ef,Nicmorus:2008vb}).
        Irrespective of the choice of $\Lambda$, hadronic ground-state properties have turned out to be insensitive
        to the value of $\eta$ in a certain range \cite{Maris:1999nt,Krassnigg:2009zh}. Consequently, with
        Eqs.~(\ref{EichmannG_RL:kernel}--\ref{EichmannG_couplingMT}) and $\Lambda$, the input of the Faddeev equation is completely specified
        with all parameters already fixed to meson properties.

%%%%%%%%%%%%%%%%%%%%%%%%%%%%%%%%%%%%%%%%%%%%%%%%%%%%%%%%%%%%%%%%%%%%%%%%%%%%%%%%%%%%%%%%%%%%%%%%%%%%%%%%%%%%%%%%%%%%%%%%%%%%%%%%%%%%%%%%%%%%%%%%%%%%%%%%%%%%%%%%%%%%%%%%%%%%%%%
%%%%%%%%%%%%%%%%%%%%%%%%%%%%%%%%%%%%%%%%%%%%%%%%%%%%%%%%%%%%%%%%%%%%%%%%%%%%%%%%%%%%%%%%%%%%%%%%%%%%%%%%%%%%%%%%%%%%%%%%%%%%%%%%%%%%%%%%%%%%%%%%%%%%%%%%%%%%%%%%%%%%%%%%%%%%%%%
%%%%%%%%%%%%%%%%%%%%%%%%%%%%%%%%%%%%%%%%%%%%%%%%%%%%%%%%%%%%%%%%%%%%%%%%%%%%%%%%%%%%%%%%%%%%%%%%%%%%%%%%%%%%%%%%%%%%%%%%%%%%%%%%%%%%%%%%%%%%%%%%%%%%%%%%%%%%%%%%%%%%%%%%%%%%%%%
%%%%%%%%%%%%%%%%%%%%%%%%%%%%%%%%%%%%%%%%%%%%%%%%%%%%%%%%%%%%%%%%%%%%%%%%%%%%%%%%%%%%%%%%%%%%%%%%%%%%%%%%%%%%%%%%%%%%%%%%%%%%%%%%%%%%%%%%%%%%%%%%%%%%%%%%%%%%%%%%%%%%%%%%%%%%%%%

     \section{Structure of the nucleon amplitude} \label{EichmannG_sec:nucleonamp}

            With the kernel $\widetilde{K}$ of the Faddeev equation \eqref{EichmannG_faddeev:eq} determined, we still need to find
            expressions for the basis elements which constitute the nucleon amplitude according to Eq.\,\eqref{EichmannG_faddeev:amp}.
            A general spinor four-point function which depends on 3 independent momenta involves 128 components of positive parity.
            These can be conveniently expressed through tensor products of two Dirac matrices, for which
             we adopt the following notation:
             \begin{eqnarray*}
                 (A\otimes B)_{\alpha\beta,\gamma\delta} &=& A_{\alpha\beta} B_{\gamma\delta}\,, \\
                 (A_1\otimes B_1) (A_2\otimes B_2) &=& (A_1 A_2)\otimes (B_1 B_2)\,, \\
                 (A \otimes B )^T &=& A^T \otimes B^T \,, \\
                 \text{Tr} \left\{ A\otimes B\right\} &=& \text{Tr}\,\{A\} \, \text{Tr}\,\{B\}\,.
             \end{eqnarray*}
            We define a charge-conjugated four-point function via
            \begin{equation}
            \begin{split}
                 & \overline{\left(A\otimes B\right)}\,(p,q,P) := \\
                 &= (C \otimes C) (A\otimes B)^T(-p,-q,-P) (C\otimes C)^T = \\
                 &= \overline{A}(p,q,P) \otimes \overline{B}(p,q,P)\,,
            \end{split}
            \end{equation}
            where $C=\gamma^4 \gamma^2$ is the charge-conjugation matrix.
            A further classification into the 64-dimensional subspaces of po\-sitive and negative energy,
            describing the nucleon's $(1/2)^+$ and $\overline{(1/2)^+}$ states, is obtained by
            attaching the respective projectors $\Lambda^\pm(P) = (\mathds{1}\pm \hat{\Slash{P}})/2$
            to the nucleon leg in the tensor product: $A\otimes B \,\Lambda^\pm$. % and $A\otimes B \,\Lambda^-$.
            The negative-parity states $(1/2)^-$ and $\overline{(1/2)^-}$ constitute another 128-dimensional vector space
            generated from basis elements of the type $A\gamma^5\otimes B \,\Lambda^\pm$.

            To construct a suitable basis for the positive-parity and positive-energy nucleon, we start from the expressions
             \begin{equation}
                 \Omega^{r}(P) = \Lambda^r(P) \,\gamma_5 C \otimes \Lambda^+(P)\,,
             \end{equation}
            with $r=\pm$ according to the projectors $\Lambda^r(P)$.
            In the quark-diquark model,
            $\Omega^{+} + \Omega^{-} = \gamma_5 C \otimes \Lambda^+$ is related to the dominant scalar-diquark part in the nucleon amplitude.
            To include the momentum dependence of the amplitude,
            it is convenient to choose a set of momenta $\{\widehat{p_T}, \widehat{q_{t}}, \widehat{P}\}$ which are orthonormal with respect to the Euclidean metric, i.e.
             \begin{equation}\label{EichmannG_orth-momenta}
             \begin{split}
                 &\widehat{p_T}^2 = \widehat{q_{t}}^2 = \widehat{P}^2 = 1, \\
                 &\widehat{p_T}\cdot \widehat{q_{t}} = \widehat{p_T}\cdot\widehat{P} = \widehat{q_{t}}\cdot\widehat{P} = 0\,.
             \end{split}
             \end{equation}
             This can be realized via
             \begin{equation}\label{EichmannG_orth-momenta2}
                 p_T^\mu := T^{\mu\nu}_P  \,p^\nu,  \qquad
                 q_{t}^\mu := T^{\mu\nu}_{p_T} \,T^{\nu\lambda}_{P} \,q^\lambda = T^{\mu\nu}_{{p_T}} \,q_T^\nu\,.
             \end{equation}
             Consider now the four quantities %covariants $\Gamma_i(p,q,P)$ defined by
             \begin{equation}\label{EichmannG_FADEEV:Gamma_i}
                \Gamma_i(p,q,P) \in \left\{  \mathds{1},\;
                                                       \textstyle\frac{1}{2}\,\displaystyle[ \widehat{\Slash{p}_T}, \widehat{\Slash{q}_{t}} ],\;
                                                       \widehat{\Slash{p}_T},\;
                                                       \widehat{\Slash{q}_{t}}
                                              \right\}\,
             \end{equation}
             where, because of Eq.\,\eqref{EichmannG_orth-momenta}, one has $\textstyle\frac{1}{2}\,\displaystyle[ \widehat{\Slash{p}_T}, \widehat{\Slash{q}_{t}} ] =  \widehat{\Slash{p}_T} \,\widehat{\Slash{q}_{t}}$.
             The linear span of the 8 matrices $\Gamma_i(p,q,P)\,\Lambda^r(P)$
             equals the linear span of the basis elements % constructed from $\mathcal{S} \,\gamma_5 C \otimes \,\mathcal{S}\,\Lambda^+$,    where $\mathcal{S}$ denotes the set
             \begin{equation} \label{EichmannG_basis:general}
                 \left\{ \, \mathds{1}, \, \Slash{p}\,, \,\Slash{q}\,, \,\Slash{P}\,, \,\Slash{p}\,\Slash{P}\,, \,\Slash{q}\,\Slash{P}\,, \,\Slash{p}\,\Slash{q}\,, \,\Slash{p}\,\Slash{q}\,\Slash{P} \,\,  \right\} ,
             \end{equation}
             whereas $\Gamma_i(p,q,P)\,\Lambda^+(P)$ reduces this set to
             \begin{equation} \label{EichmannG_basis:general2}
                  \left\{ \, \mathds{1}, \, \Slash{p}\,, \,\Slash{q}\,, \,\Slash{p}\,\Slash{q} \,  \right\}\Lambda^+
             \end{equation}
             since $\widehat{\Slash{P}}\,\Lambda^+ = \Lambda^+$.
             Hence the quantities $(\Gamma_i   \otimes  \Gamma_j ) \, \Omega^{r}$ define 32 linearly independent basis tensors.

    \renewcommand{\arraystretch}{1.2}

             A complete orthogonal basis for the nucleon amplitude
             is given by the 64-dimensional set $\{\mathsf{S}_{ij}^r,\,\mathsf{P}_{ij}^r\}$ defined by
            \begin{equation} \label{EichmannG_basisSP}
                \left( \begin{array}{c} \mathsf{S}_{ij}^r(p,q,P) \\ \mathsf{P}_{ij}^r(p,q,P) \end{array}\right) :=
                \left( \begin{array}{c} \mathds{1}\otimes\mathds{1} \\ \gamma^5 \otimes \gamma^5 \end{array}\right) (\Gamma_i   \otimes  \Gamma_j ) \, \Omega^{r} ,
            \end{equation}
             with $i,j=1\dots 4$.
             The symbols $\mathsf{S}$ and $\mathsf{P}$ were chosen to reflect a combination of two scalar or pseudoscalar covariants
             whose product again exhibits positive parity.
             All possible further basis elements, e.g.
            \begin{equation} \label{EichmannG_basisAVT}
                \left( \begin{array}{c} \mathsf{A}_{ij}^r(p,q,P) \\ \mathsf{V}_{ij}^r(p,q,P) \\ \mathsf{T}_{ij}^r(p,q,P) \end{array}\right) :=
                \left( \begin{array}{c} \gamma^\mu_T \,\gamma_5 \otimes \gamma^\mu_T\,\gamma_5 \\ \gamma^\mu_T  \otimes \gamma^\mu_T \\ \sigma^{\mu\nu}_T \otimes   \sigma^{\mu\nu}_T  \end{array}\right)
                        (\Gamma_i   \otimes  \Gamma_j ) \, \Omega^{r} ,
            \end{equation}
             where $\gamma_T^\mu = T_P^{\mu\nu} \gamma^\nu$ and $\sigma_T^{\mu\nu} = -i\,[\gamma_T^\mu,\gamma_T^\nu]/2$,
             linearly depend on those in Eq.\,\eqref{EichmannG_basisSP}; respective relations are given in Table~\ref{EichmannG_tab:faddeev:basis-tf}. For instance,
             the dominant contributions to the Faddeev amplitude are the elements
             \begin{equation*}
             \begin{split}
                 \gamma_5 C \otimes \Lambda^+ &= \sum_{r=\pm} \mathsf{S}_{11}^r \,,\\
                 \gamma^\mu_T C \otimes \gamma^\mu_T \gamma_5 \Lambda^+ &= \sum_{r=\pm} \mathsf{A}_{11}^r = \sum_{r=\pm} \left( r \,\mathsf{S}_{22}^r + \mathsf{P}_{33}^r + \mathsf{P}_{44}^r \right).
             \end{split}
             \end{equation*}
             In the quark-diquark model, these correspond to scalar-scalar and axialvector-axialvector combinations of diquark and quark-diquark amplitudes
             for either of the three diagrams appearing in the Faddeev equation in Fig.~\ref{EichmannG_fig:faddeev}.

    \renewcommand{\arraystretch}{1.5}

             \begin{table*}[p]

               \caption{   Relations between the basis elements $\mathsf{A}_{ij}$ of Eq.\,\eqref{EichmannG_basisAVT} and $\{\mathsf{S}_{ij},\,\mathsf{P}_{ij}\}$.
                           The corresponding relations for $\mathsf{V}_{ij}$ are obtained by interchanging $\mathsf{S}_{ij}\leftrightarrow \mathsf{P}_{ij}$.
                           Similar dependencies hold for the $\mathsf{T}_{ij}$, e.g.: $\mathsf{T}_{11}^+ = -2\, \mathsf{A}_{11}^+$.
                           The superscripts $r=\pm$ are not displayed for better readability. }\label{EichmannG_tab:faddeev:basis-tf}

                \begin{center}
                \begin{tabular}{    @{\quad} l @{\quad\quad\quad}  l @{\quad\quad\quad} l @{\quad\quad\quad}  l  @{\quad}  } \hline %\rule{-1mm}{2.15cm}

                         &&& \\[-0.45cm]

                         $\mathsf{A}_{11} = \mathsf{P}_{33} + \mathsf{P}_{44} + r \mathsf{S}_{22}$ &
                         $\mathsf{A}_{12} = \mathsf{P}_{34} - \mathsf{P}_{43} - r \mathsf{S}_{21}$ &
                         $\mathsf{A}_{13} = \mathsf{P}_{31} - \mathsf{P}_{42} + r \mathsf{S}_{24}$ &
                         $\mathsf{A}_{14} = \mathsf{P}_{32} + \mathsf{P}_{41} - r \mathsf{S}_{23}$ \\

                         $\mathsf{A}_{21} = \mathsf{P}_{43} - \mathsf{P}_{34} - r \mathsf{S}_{12}$ &
                         $\mathsf{A}_{22} = \mathsf{P}_{33} + \mathsf{P}_{44} + r \mathsf{S}_{11}$ &
                         $\mathsf{A}_{23} = \mathsf{P}_{32} + \mathsf{P}_{41} - r \mathsf{S}_{14}$ &
                         $\mathsf{A}_{24} = \mathsf{P}_{42} - \mathsf{P}_{31} + r \mathsf{S}_{13}$ \\

                         $\mathsf{A}_{31} = \mathsf{P}_{13} - \mathsf{P}_{24} + r \mathsf{S}_{42}$ &
                         $\mathsf{A}_{32} = \mathsf{P}_{23} + \mathsf{P}_{14} - r \mathsf{S}_{41}$ &
                         $\mathsf{A}_{33} = \mathsf{P}_{11} + \mathsf{P}_{22} + r \mathsf{S}_{44}$ &
                         $\mathsf{A}_{34} = \mathsf{P}_{12} - \mathsf{P}_{21} - r \mathsf{S}_{43}$ \\

                         $\mathsf{A}_{41} = \mathsf{P}_{14} + \mathsf{P}_{23} - r \mathsf{S}_{32}$ &
                         $\mathsf{A}_{42} = \mathsf{P}_{24} - \mathsf{P}_{13} + r \mathsf{S}_{31}$ &
                         $\mathsf{A}_{43} = \mathsf{P}_{21} - \mathsf{P}_{12} - r \mathsf{S}_{34}$ &
                         $\mathsf{A}_{44} = \mathsf{P}_{11} + \mathsf{P}_{22} + r \mathsf{S}_{33}$ \\[0.15cm] \hline

                \end{tabular}
                \end{center}

        \end{table*}

    \renewcommand{\arraystretch}{1.5}

             \begin{table*}[p]

               \caption{Orthonormal Dirac basis $\mathsf{X}_{ij,k}^r$ of Eq.\,\eqref{EichmannG_basisX} constructed from a partial-wave decomposition.
                                       The first two columns denote the eigenvalues of total quark spin $s$ and intrinsic orbital angular momentum $l$ in the nucleon rest frame.
                                       The third and fourth columns define the relation between the $\mathsf{X}_{ij}^r$ and the basis elements
                                       from Eq.\,(\ref{EichmannG_basisSP}-\ref{EichmannG_basisAVT}).
                                       Each row involves 4 covariants; the superscripts $r=\pm$ are not displayed for better readability.
                                       The fifth column shows the momentum-dependent covariants $T_{ij}$ which appear in Eq.\,\eqref{EichmannG_basisX};
                                       we have abbreviated $\widehat{p_T} \rightarrow p$ and $\widehat{q_t} \rightarrow q$ for clarity.
                                       }\label{EichmannG_tab:faddeev:basis}

                \begin{center}
                \begin{tabular}{   | c | c || c   |    c      ||  @{\quad}l@{\quad} | } \hline \rule{0mm}{0.5cm}

                 $s$                 &  $l$  &   $\mathsf{X}^r_{1j,1}$ &
                                                 $\mathsf{X}^r_{1j,2}$ &
                                                 $T_{1j}$
                                                 \\ [0.15cm]  \hline\hline

                 $\nicefrac{1}{2}$   &  $0$  &   $\quad\;\;\mathsf{S}_{11}\quad\;\;$    &
                                                 $\quad\;\;\mathsf{P}_{11}\quad\;\;$    &
                                                 $\mathds{1}\otimes \mathds{1}$
                                                 \\

                 $\nicefrac{1}{2}$   &  $1$  &   $\mathsf{S}_{12}$     &
                                                 $\mathsf{P}_{12}$     &
                                                 $\mathds{1}\otimes \textstyle\frac{1}{2}\, [\, \Slash{p}, \Slash{q} \,]$
                                                 \\

                 $\nicefrac{1}{2}$   &  $1$  &   $\mathsf{S}_{13}$     &
                                                 $\mathsf{P}_{13}$     &
                                                 $\mathds{1}\otimes \Slash{p}$
                                                 \\

                 $\nicefrac{1}{2}$   &  $1$  &   $\mathsf{S}_{14}$     &
                                                 $\mathsf{P}_{14}$     &
                                                 $\mathds{1}\otimes \Slash{q}$
                                                 \\ [0.1cm]  \hline\hline \rule{0mm}{0.5cm}

                 $s$                 &  $l$  &   $\sqrt{3}\,\,\mathsf{X}^r_{2j,1}$   &
                                                 $\sqrt{3}\,\,\mathsf{X}^r_{2j,2}$  &
                                                 $\sqrt{3}\,\,T_{2j}$
                                                 \\ [0.15cm]  \hline\hline

                 $\nicefrac{1}{2}$   &  $0$  &   $\quad\;\;\mathsf{V}_{11}\quad\;\;$    &
                                                 $\quad\;\;\mathsf{A}_{11}\quad\;\;$    &
                                                 $\gamma^\mu_T  \otimes \gamma^\mu_T $
                                                 \\

                 $\nicefrac{1}{2}$   &  $1$  &   $\mathsf{V}_{12}$     &
                                                 $\mathsf{A}_{12}$     &
                                                 $\gamma^\mu_T  \otimes \gamma^\mu_T  \,\textstyle\frac{1}{2}\, [\, \Slash{p}, \Slash{q} \,]$
                                                 \\

                 $\nicefrac{1}{2}$   &  $1$  &   $\mathsf{V}_{13}$     &
                                                 $\mathsf{A}_{13}$     &
                                                 $\gamma^\mu_T  \otimes \gamma^\mu_T  \,\Slash{p}$
                                                 \\

                 $\nicefrac{1}{2}$   &  $1$  &   $\mathsf{V}_{14}$     &
                                                 $\mathsf{A}_{14}$     &
                                                 $\gamma^\mu_T  \otimes \gamma^\mu_T \,\Slash{q}$
                                                 \\ [0.1cm]  \hline\hline \rule{0mm}{0.5cm}

                 $s$                 &  $l$  &   $\sqrt{6}\,\,\mathsf{X}^r_{3j,1}$
                                             &   $\sqrt{6}\,\,\mathsf{X}^r_{3j,2}$
                                             & $\sqrt{6}\,\,T_{3j}$
                                             \\ [0.15cm]  \hline\hline

                 $\nicefrac{3}{2}$   &  $2$  &   $3\,\mathsf{S}_{33}-\mathsf{V}_{11}  $
                                             &   $3\,\mathsf{P}_{33}-\mathsf{A}_{11}$
                                             &   $3\,\Slash{p} \otimes \Slash{p} - \gamma^\mu_T \otimes \gamma^\mu_T$
                                             \\

                 $\nicefrac{3}{2}$   &  $1$  &   $\quad 3\,\mathsf{S}_{34}-3\,\mathsf{S}_{43}-2\,\mathsf{V}_{12} \quad $
                                             &   $\quad 3\,\mathsf{P}_{34}-3\,\mathsf{P}_{43}-2\,\mathsf{A}_{12} \quad $
                                             &   $3\left( \Slash{p} \otimes \Slash{q} -\Slash{q} \otimes \Slash{p} \right) - \gamma^\mu_T \otimes \gamma^\mu_T \,[\, \Slash{p}, \Slash{q} \,]$
                                             \\

                 $\nicefrac{3}{2}$   &  $1$  &   $3\,\mathsf{S}_{31}-\mathsf{V}_{13}$
                                             &   $3\,\mathsf{P}_{31}-\mathsf{A}_{13}$
                                             &   $3\,\Slash{p} \otimes \mathds{1} - \gamma^\mu_T \otimes \gamma^\mu_T \, \Slash{p}$
                                             \\

                 $\nicefrac{3}{2}$   &  $1$  &   $3\,\mathsf{S}_{41}-\mathsf{V}_{14}$
                                             &   $3\,\mathsf{P}_{41}-\mathsf{A}_{14}$
                                             &   $3\,\Slash{q} \otimes \mathds{1} - \gamma^\mu_T \otimes \gamma^\mu_T \, \Slash{q}$
                                             \\ [0.1cm]  \hline  \hline \rule{0mm}{0.5cm}

                 $s$                 &  $l$  &   $\sqrt{2}\,\,\mathsf{X}^r_{4j,1}$
                                             &   $\sqrt{2}\,\,\mathsf{X}^r_{4j,2}$
                                             &   $\sqrt{2}\,\,T_{4j}$
                                             \\ [0.15cm]  \hline   \hline

               $\quad\nicefrac{3}{2}\quad$ &  $\quad 2 \quad$  &   $2\,\mathsf{S}_{44}+\mathsf{S}_{33}-\mathsf{V}_{11}$
                                                     &   $2\,\mathsf{P}_{44}+\mathsf{P}_{33}-\mathsf{A}_{11}$
                                                     &   $\Slash{p} \otimes \Slash{p} + 2\,\Slash{q} \otimes \Slash{q} -  \gamma^\mu_T \otimes \gamma^\mu_T$
                                                     \\

                 $\nicefrac{3}{2}$   &  $2$          &   $\mathsf{S}_{34}+\mathsf{S}_{43}$
                                                     &   $\mathsf{P}_{34}+\mathsf{P}_{43}$
                                                     &   $\Slash{p} \otimes \Slash{q} + \Slash{q} \otimes \Slash{p}$
                                                     \\

                 $\nicefrac{3}{2}$   &  $2$          &   $-2\,\mathsf{S}_{42}+\mathsf{S}_{31}-\mathsf{V}_{13}$
                                                     &   $-2\,\mathsf{P}_{42}+\mathsf{P}_{31}-\mathsf{A}_{13}$
                                                     &   $\Slash{q} \otimes [\, \Slash{q}, \Slash{p} \,] - \textstyle\frac{1}{2}\,\gamma^\mu_T \otimes [\,\gamma^\mu_T, \, \Slash{p}\,]$
                                                     \\

                 $\nicefrac{3}{2}$   &  $2$          &   $2\,\mathsf{S}_{32}+\mathsf{S}_{41}-\mathsf{V}_{14}$
                                                     &   $2\,\mathsf{P}_{32}+\mathsf{P}_{41}-\mathsf{A}_{14}$
                                                     &   $\Slash{p} \otimes [\, \Slash{p}, \Slash{q} \,] - \textstyle\frac{1}{2}\,\gamma^\mu_T \otimes [\,\gamma^\mu_T ,\, \Slash{q}\,]$
                                                     \\ [0.1cm]  \hline

                \end{tabular}
                \end{center}

        \end{table*}

             \begin{table*}[p]

               \caption{  Irreducible multiplets of the permutation group $\mathbb{S}_3$, constructed from the 8 covariants
                          $\{\mathsf{S}_{11}^r,\,\mathsf{P}_{11}^r\,,\mathsf{A}_{11}^r,\,\mathsf{V}_{11}^r\}$. }\label{EichmannG_tab:faddeev:s3rep}

                \begin{center}
                \begin{tabular}{    @{\quad} l @{\quad\quad\quad}  l @{\quad\quad\quad} l @{\quad\quad\quad}  l  @{\quad}  } \hline %\rule{-1mm}{2.15cm}

                         &&& \\[-0.45cm]

                         $\psi_\mathcal{M_A}^1 = \mathsf{S}_{11}^+$ &
                         $\psi_\mathcal{M_A}^2 = \sum_r \mathsf{P}_{11}^r + \mathsf{S}_{11}^-$ &
                         $\psi_\mathcal{M_A}^3 = \sum_r \left( \mathsf{V}_{11}^r - \mathsf{P}_{11}^r\right) + 2\,\mathsf{S}_{11}^-$ &
                         $\psi_\mathcal{A} =  \sum_r \left( \mathsf{V}_{11}^r + \mathsf{P}_{11}^r\right) - 2\,\mathsf{S}_{11}^-$ \\

                         $\psi_\mathcal{M_S}^1 = \mathsf{A}_{11}^+$ &
                         $\psi_\mathcal{M_S}^2 = \sum_r r \mathsf{V}_{11}^r - \mathsf{A}_{11}^-$ &
                         $\psi_\mathcal{M_S}^3 = \sum_r r \left( \mathsf{V}_{11}^r +3 \mathsf{P}_{11}^r\right) + 2\,\mathsf{A}_{11}^-$ &
                         $\psi_\mathcal{S} =  \sum_r r \left( -\mathsf{V}_{11}^r +3 \mathsf{P}_{11}^r\right)  - 2\,\mathsf{A}_{11}^-$ \\[0.15cm] \hline

                \end{tabular}
                \end{center}

        \end{table*}

    \renewcommand{\arraystretch}{1.2}

             A partial-wave decomposition (see Table~\ref{EichmannG_tab:faddeev:basis} and App.~\ref{EichmannG_sec:partialwave}) leads to linear combinations of the $\{\mathsf{S}_{ij}^r,\,\mathsf{P}_{ij}^r\}$
             as eigenstates of quark-spin and orbital angular momentum operators $\vect{S}^2$ and $\vect{L}^2$
             in the nucleon rest frame. The 64 basis covariants (32 each for total quark spin $s=1/2$ and $s=3/2$, respectively)
             can be arranged into sets of 8 $s$-waves ($l=0$), 36 $p$-waves ($l=1$), and 20 $d$-waves ($l=2$) which we denote
             collectively by
            \begin{equation} \label{EichmannG_basisX}
                \left( \begin{array}{cc} \mathsf{X}_{ij,1}^r \\ \mathsf{X}_{ij,2}^r \end{array}\right) :=
                \left( \begin{array}{cc} \mathds{1}\otimes\mathds{1} \\ \gamma^5 \otimes \gamma^5 \end{array}\right) T_{ij} \, \Omega^{r} \,,
            \end{equation}
            with $i,j=1\dots 4$ and $r=\pm$.
            The $T_{ij}$ carry the relative-momentum dependence and are defined in Table~\ref{EichmannG_tab:faddeev:basis}.
            For instance, the eigenstates of $s=\nicefrac{1}{2}$ correspond to $i=1,2$ and read
            \begin{equation}
            \begin{split}
                T_{1j} &= \Gamma_1 \otimes \Gamma_j\,,\\
                T_{2j} &= \textstyle{\frac{1}{\sqrt{3}}}\,(\gamma^\mu_T  \otimes \gamma^\mu_T ) \, (\Gamma_1 \otimes \Gamma_j)\,,
            \end{split}
            \end{equation}
            whereas the $s=\nicefrac{3}{2}$ eigenstates ($i=3,4$) are more complicated.
            The relations between the $\mathsf{X}_{ij,k}^r$ and the basis elements $\{\mathsf{S}_{ij}^r,\,\mathsf{P}_{ij}^r\,,\mathsf{A}_{ij}^r,\,\mathsf{V}_{ij}^r\}$
            are stated in Table~\ref{EichmannG_tab:faddeev:basis} as well.

             \renewcommand{\arraystretch}{1}

            As illustrated in App.~\ref{EichmannG_sec:threespinor}, the basis elements $\mathsf{X}_{ij}^r$ can be expressed in terms of quark three-spinors frequently used in the literature, e.g. Ref.\,\cite{Carimalo:1992ia}.
            Moreover, they satisfy the following orthogonality relation:
                 \begin{equation}\label{EichmannG_faddeev:orthogonality}
                     \textstyle\frac{1}{4} \,\displaystyle \text{Tr}\,\{\conjg{\mathsf{X}}^r_{ij}\,\mathsf{X}^{r'}_{i'j'}\} =
                     \textstyle\frac{1}{4} \,\displaystyle \big(\conjg{\mathsf{X}}^r_{ij}\big)_{\beta\alpha,\delta\gamma}\,\big(\mathsf{X}^{r'}_{i'j'}\big)_{\alpha\beta,\gamma\delta} =
                      \delta_{ii'}\,\delta_{jj'}\,\delta_{rr'}\,.
                 \end{equation}
             As an example, consider the scalar-scalar combinations:
                 \begin{equation*}
                 \begin{split}
                     \conjg{\mathsf{S}}_{ij}^r(p,q,P) =\,& C\,\big\{\Gamma_i(-p,-q,-P)\, \Lambda^r(-P) \, \gamma_5 C\big\}^T C^T    \\
                                                         & \otimes C\,\big\{\Gamma_j(-p,-q,-P)\, \Lambda^+(-P)\big\}^T C^T = \\
                                                      =\,& (C^T\!\gamma_5)\,(\Lambda^r \conjg{\Gamma}_i ) \,\otimes  (\Lambda^+\conjg{\Gamma}_j ),
                 \end{split}
                 \end{equation*}
            where we used the relations $\conjg{\Lambda}^r(P) = C\,\Lambda^r(-P)^T\,C^T=\Lambda^r(P)$ and $\conjg{\Gamma}_i(p,q,P) = C\,\Gamma_i(-p,-q,-P)^T\,C^T$.
            As a result one obtains
                 \begin{equation*}
                 \begin{split}
                     \textstyle\frac{1}{4} \,\displaystyle \text{Tr}\{\conjg{\mathsf{S}}^r_{ij}\,\mathsf{S}^{r'}_{i'j'}\} =&
                     \textstyle\frac{1}{4} \,\displaystyle \text{Tr}\{ \conjg{\Gamma}_i \,\Gamma_{i'}\,\Lambda^{r'}\Lambda^r \} \,\text{Tr}\{ \conjg{\Gamma}_j\,\Gamma_{j'}\,\Lambda^+ \}  = \\
                     =& \delta_{ii'}\,\delta_{jj'}\,\delta_{rr'}\,,
                 \end{split}
                 \end{equation*}
            since $\Lambda^{r'}\Lambda^r = \delta_{rr'}\Lambda^r$ and $\text{Tr} \left\{ \conjg{\Gamma}_i \,\Gamma_{i'}\,\Lambda^r \right\} = 2\,\delta_{ii'}$.

            \begin{figure}[!t]
                    \begin{center}
                    \includegraphics[scale=0.345]{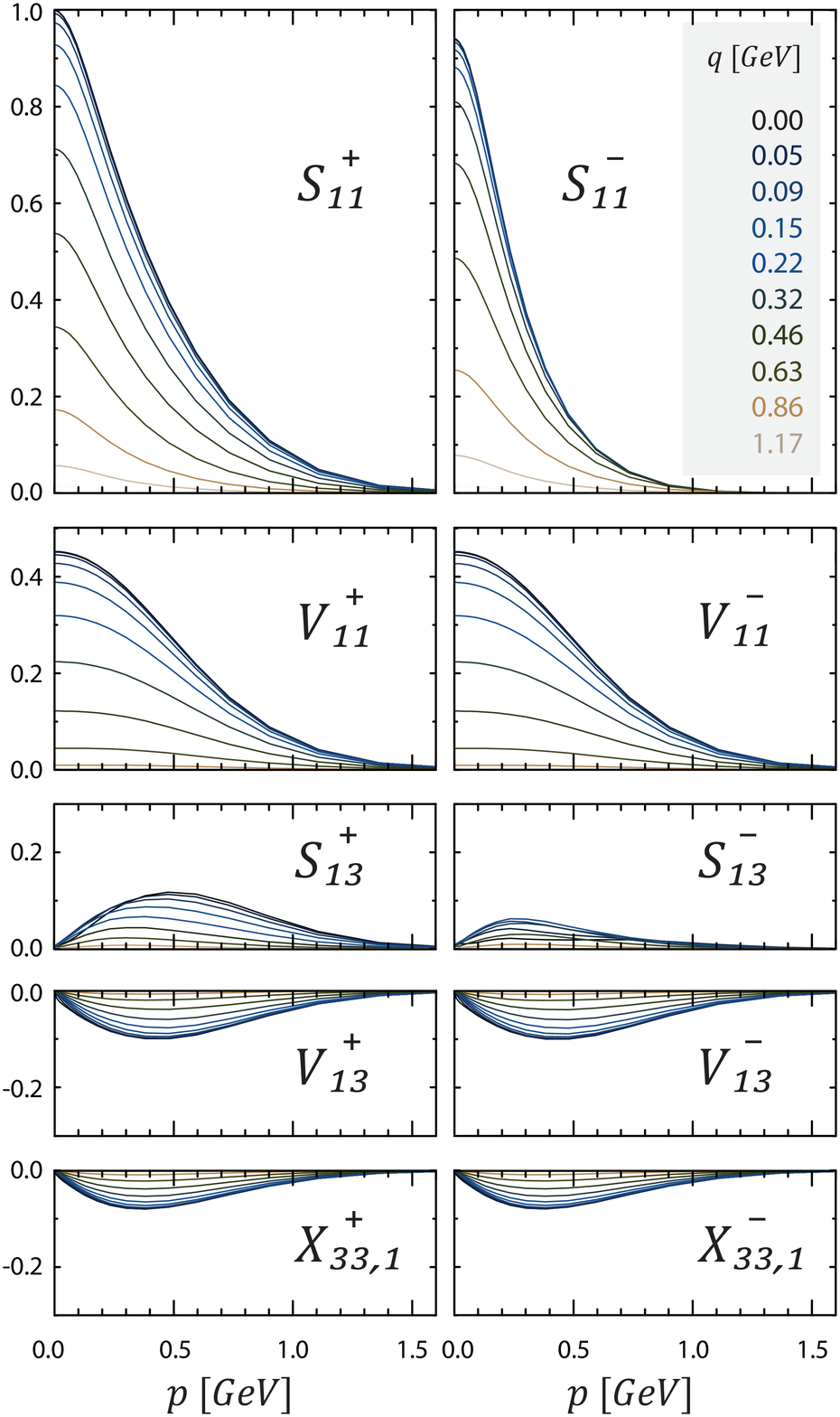}
                    \caption{  Zeroth Chebyshev moments of the dressing functions corresponding to the
                               dominant covariants in the Faddeev amplitude $\Psi_\mathcal{M_A}$, plotted as a function of $\sqrt{p^2}$ and $\sqrt{q^2}$.
                               The various curves represent the falloff in the variable $\sqrt{q^2}$.
                              } \label{EichmannG_fig:amplitudes}
                    \vspace{0.5cm}
                    \includegraphics[scale=0.345]{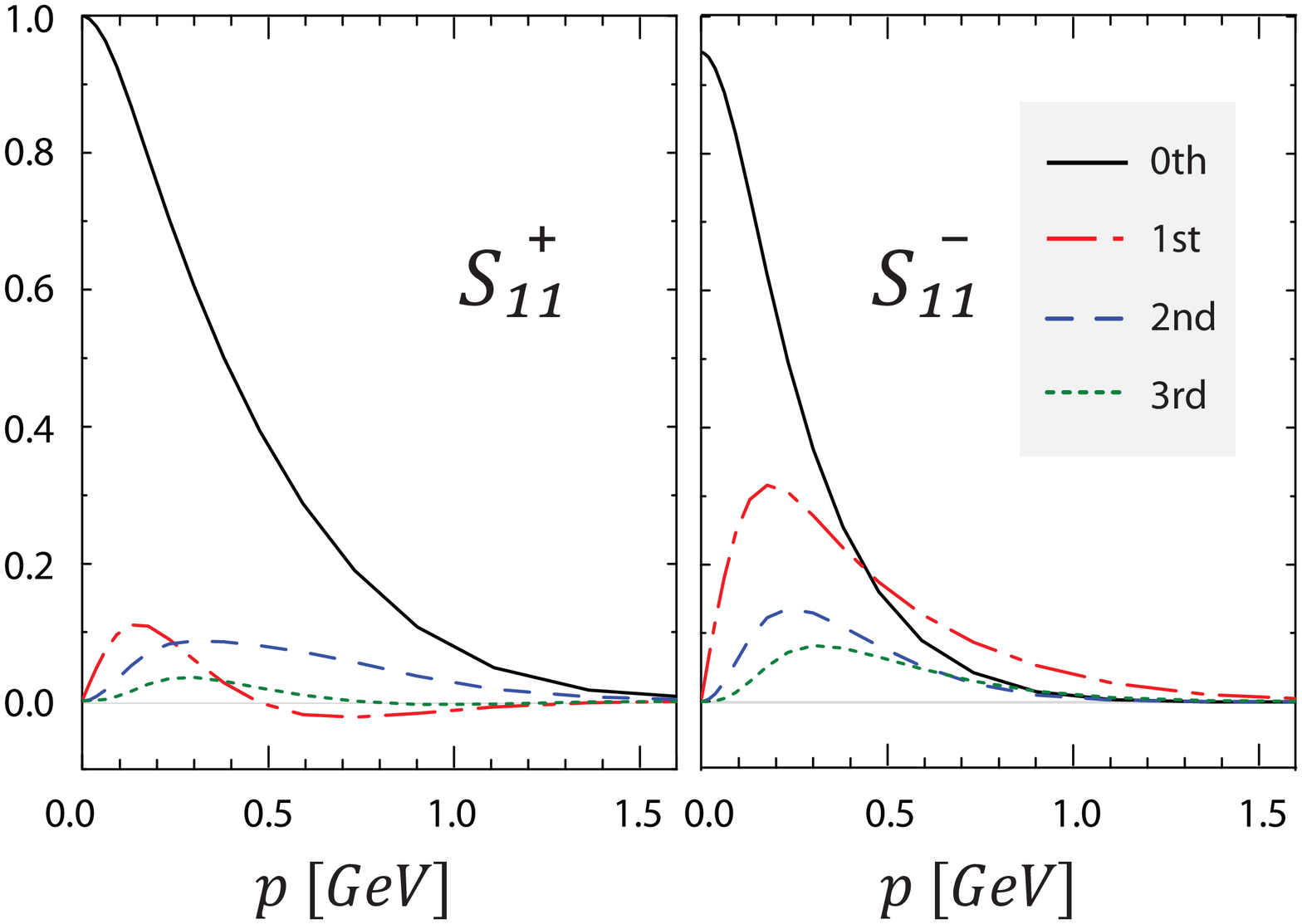}
                    \caption{  First four Chebyshev moments in the variable $z_1$ of the dressing functions
                               associated with the amplitudes $\mathsf{S}_{11}^r$, evaluated at $q^2=0$. }
                    \end{center} \label{EichmannG_fig:3}
            \end{figure}

             The Pauli principle requires the Faddeev amplitude to be antisymmetric under exchange of any two quarks.
             The Faddeev kernel $\widetilde{K}_{(3)}$ is invariant
             under the permutation group $\mathbb{S}_3$. The eigenstates of the Faddeev kernel
             can hence be arranged into irreducible $\mathbb{S}_3$ multiplets
             \begin{equation}\label{EichmannG_psi:irreps}
                 \Psi_\mathcal{S}, \; \Psi_\mathcal{A}, \; \left(\begin{array}{c} \Psi_\mathcal{M_A} \\ \Psi_\mathcal{M_S} \end{array}\right),
             \end{equation}
             of which the first two (totally symmetric or antisymmetric) solutions are unphysical while the mixed-symmetry doublet constitutes the Dirac part of the nucleon amplitude.
             Taking into account the flavor and color structure, the full Dirac--flavor--color amplitude reads
              \begin{equation}\label{EichmannG_FE:nucleon_amplitude_full-2}
                    \Psi(p,q,P) =  \Big\{ \Psi_\mathcal{M_A}  \mathsf{T}_\mathcal{M_A} + \Psi_\mathcal{M_S}  \mathsf{T}_\mathcal{M_S} \Big\} \frac{\varepsilon_{ABC}}{\sqrt{6}} \,,
              \end{equation}
             where $\varepsilon_{ABC}$ is the antisymmetric color-singlet wave function and
              \begin{equation} \label{EichmannG_FAD:flavor1}
                  \mathsf{T}_\mathcal{M_A} = \textstyle\frac{1}{\sqrt{2}}\,i \sigma_2\otimes \mathds{1} \,, \quad
                  \mathsf{T}_\mathcal{M_S} = -\textstyle\frac{1}{\sqrt{6}}\, \vect{\sigma}\,i\sigma_2 \otimes \vect{\sigma}
              \end{equation}
             denote the isospin-$1/2$ flavor tensors involving the Pauli matrices $\sigma_i$.
             A projection onto the proton or neutron flavor states involves a contraction of the rearmost
             flavor index with either of the two isospin basis states $(1,0)$ or $(0,1)$.

             A flavor-dependent kernel in the Faddeev equation will mix $\Psi_\mathcal{M_A}$ and $\Psi_\mathcal{M_S}$ whose dominant contributions
             are given by $\mathsf{S}_{11}^+$ and $\mathsf{A}_{11}^+$, respectively.
             However, since the rain\-bow-ladder kernel presently employed is flavor-independent and we consider only equal quark masses,
             the equations for the Dirac amplitudes $\Psi_\mathcal{M_A}$ and $\Psi_\mathcal{M_S}$ in
             Eq.\,\eqref{EichmannG_FE:nucleon_amplitude_full-2} decouple because of the orthogonality of
             the two flavor tensors $\mathsf{T}_\mathcal{M_A}$ and $\mathsf{T}_\mathcal{M_S}$:
             \begin{equation}\label{EichmannG_FADDEEV:MS,MA:Independent}
                 \Psi = \widetilde{K}_{(3)}\,\Psi \quad \longrightarrow \quad \begin{array}{c} \Psi_\mathcal{M_A} = \widetilde{K}_{(3)}\,\Psi_\mathcal{M_A}, \\[0.1cm]
                                                                                    \Psi_\mathcal{M_S} = \widetilde{K}_{(3)}\,\Psi_\mathcal{M_S}.\end{array}
             \end{equation}
             Hence one obtains two degenerate solutions of the Faddeev equation, where by virtue of the iterative solution method
             the symmetry of the start function determines the symmetry of the resulting amplitude.

             Similarly to the analogous case of a diquark amplitude, the quark exchange symmetry does not
             reduce the number of Dirac covariants since the dressing functions $f_k(p^2,q^2,\{z\})$ transform under the permutation group as well.
             For example, if we restrict ourselves to the eight momentum-independent covariants of Table~\ref{EichmannG_tab:faddeev:basis} (with $s=\nicefrac{1}{2}$, $l=0$) and
              apply the $\mathbb{S}_3$ (anti-)symmetrizers, we can arrange these 8 basis elements into three mixed-symmetry doublets and one
              symmetric and one antisymmetric singlet which are shown in Table~\ref{EichmannG_tab:faddeev:s3rep}.
             If the coefficients $f_k$ were totally symmetric,
             e.g., by being constant or by depending only on certain symmetric combinations of $p^2$, $q^2$ and $\{z\}$ as derived in Ref.~\cite{Carimalo:1992ia},
             the nucleon amplitude would be a linear combination of the six $\mathcal{M_A}$ and $\mathcal{M_S}$ basis elements in Table~\ref{EichmannG_tab:faddeev:s3rep}:
             \begin{equation}
                 \Psi_\mathcal{M_A} = \sum_{k=1}^3 f_k\,\Psi^k_\mathcal{M_A}\,, \quad
                 \Psi_\mathcal{M_S} = \sum_{k=1}^3 f_k'\,\Psi^k_\mathcal{M_S}\,.
             \end{equation}
             Since the coefficients $f_k$ can appear in all symmetry representations,
             the inclusion of the remaining Dirac covariants $\psi_\mathcal{A}$ and $\psi_\mathcal{S}$ is however necessary,
             and the same reasoning holds if all 64 basis elements are implemented.

%%%%%%%%%%%%%%%%%%%%%%%%%%%%%%%%%%%%%%%%%%%%%%%%%%%%%%%%%%%%%%%%%%%%%%%%%%%%%%%%%%%%%%%%%%%%%%%%%%%%%%%%%%%%%%%%%%%%%%%%%%%%%%%%%%%%%%%%%%%%%%%%%%%%%%%%%%%%%%%%%%%%%%%%%%%%%%%
%%%%%%%%%%%%%%%%%%%%%%%%%%%%%%%%%%%%%%%%%%%%%%%%%%%%%%%%%%%%%%%%%%%%%%%%%%%%%%%%%%%%%%%%%%%%%%%%%%%%%%%%%%%%%%%%%%%%%%%%%%%%%%%%%%%%%%%%%%%%%%%%%%%%%%%%%%%%%%%%%%%%%%%%%%%%%%%
%%%%%%%%%%%%%%%%%%%%%%%%%%%%%%%%%%%%%%%%%%%%%%%%%%%%%%%%%%%%%%%%%%%%%%%%%%%%%%%%%%%%%%%%%%%%%%%%%%%%%%%%%%%%%%%%%%%%%%%%%%%%%%%%%%%%%%%%%%%%%%%%%%%%%%%%%%%%%%%%%%%%%%%%%%%%%%%
%%%%%%%%%%%%%%%%%%%%%%%%%%%%%%%%%%%%%%%%%%%%%%%%%%%%%%%%%%%%%%%%%%%%%%%%%%%%%%%%%%%%%%%%%%%%%%%%%%%%%%%%%%%%%%%%%%%%%%%%%%%%%%%%%%%%%%%%%%%%%%%%%%%%%%%%%%%%%%%%%%%%%%%%%%%%%%%

     \section{Results}\label{EichmannG_sec:FADDEEV:results}

             The explicit numerical implementation of the Faddeev equation is described in App.~\ref{EichmannG_sec:numerics}.
             The massive computational demand in solving the equation primarily comes from the
             five Lorentz-invariant momentum combinations of Eq.\,\eqref{EichmannG_mom-variables} upon which the amplitudes depend.
             In analogy to the separability assumption of the nucleon amplitude in the quark-diquark model
             we omit the dependence on the angular variable $z_0=\widehat{p_T}\cdot\widehat{q_T}$
             but solve for all 64 dressing functions $f_k(p^2, q^2,0,z_1,z_2)$.

             The resulting nucleon masses at the physical pion mass in both setups A and B are
             shown in Table~\ref{EichmannG_tab:results}.
                As a consequence of Eq.\,\eqref{EichmannG_FADDEEV:MS,MA:Independent},
                the two states $\Psi_\mathcal{M_A}$ and $\Psi_\mathcal{M_S}$ emerge as independent solutions of the Faddeev equation.
                Both separate equations produce approximately the same nucleon mass,
                where the deviation of $\sim 2\%$ is presumably a truncation artifact associated with the omission of the angle $z_0$.
             For either solution typically only a small number of covariants are relevant
             which are predominantly $s$-wave with a small $p$-wave admixture.
             The corresponding amplitudes for the mixed-antisymmetric solution are shown in Fig.~\ref{EichmannG_fig:amplitudes}.
                Comparing the relative strengths of the amplitudes allows to identify the dominant contributions:
                \begin{equation}
                \begin{split}
                  \Psi_\mathcal{M_A}: &  \sum_r \left\{ \mathsf{S}_{11}^r, \; \mathsf{V}_{11}^r, \; \mathsf{S}_{13}^r, \; \mathsf{V}_{13}^r, \; \mathsf{X}_{33,1}^r \right\}, \\
                  \Psi_\mathcal{M_S}: &  \sum_r \left\{ \mathsf{A}_{11}^r, \; r\mathsf{V}_{11}^r, \; r\mathsf{P}_{11}^r, \; r\mathsf{V}_{13}^r, \; \mathsf{X}_{33,2}^r \right\}.
                \end{split}
                \end{equation}

                Fig.~3 displays the angular dependence in the variable $z_1$
                through the first few Chebyshev moments of the amplitudes $\mathsf{S}_{11}^\pm$
                which contribute to $\Psi_\mathcal{M_A}$.
             The angular dependence in the
             variable $z_2$ is small compared to $z_1$. This is analogous to the quark-diquark model,
             where the dependence on the angle between
             the relative and total momentum of the two quarks in a diquark amplitude is weak.

\begin{figure}[t]
            \begin{center}
            \includegraphics[scale=0.85]{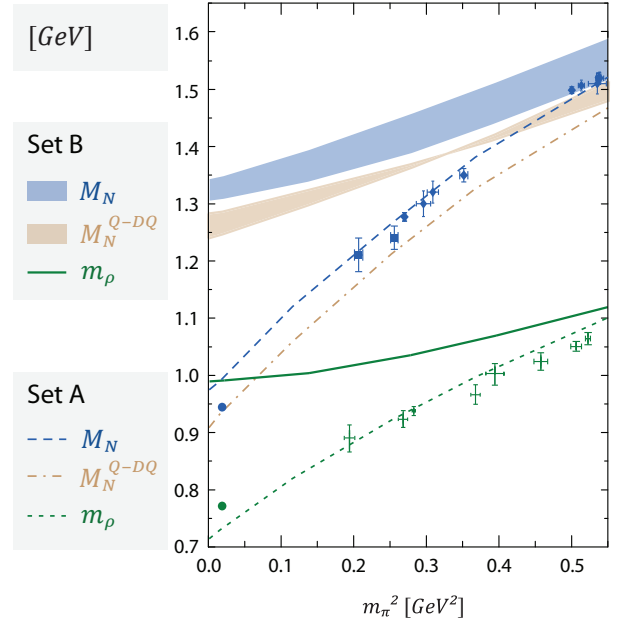}
            \caption{(adapted from Ref.~\cite{Eichmann:2009qa}) Evolution with $m_{\pi}^2$ of $m_\rho$ and $M_N$
                                     compared to lattice data (\cite{AliKhan:2003cu,Frigori:2007wa} for
                                     $M_N$ and \cite{AliKhan:2001tx,Allton:2005fb} for $m_\rho$).
                                     The quark-diquark model result for $M_N$~\cite{Eichmann:2008ef,Nicmorus:2008vb} is plotted for comparison.
                                     Dashed and dashed-dotted lines correspond to setup A;
                                     the solid line for $m_\rho$ and the bands for $M_N$ (mixed-antisymmetric
                                     solution) are the results of setup B,
                                     where the variation with $\eta$ is explicitly taken into account.
                                     Dots denote the experimental values.} \label{EichmannG_fig:FADDEEV:nucleon-mass}
            \end{center}
            \end{figure}

             The evolution of $M_N$ and the $\rho$-meson mass from the BSE vs.~$m_\pi^2$ is plotted in
	         Fig.~\ref{EichmannG_fig:FADDEEV:nucleon-mass} and compared to lattice results.
             The findings for $M_N$ are qualitatively similar to those for $m_\rho$: setup A, where the coupling strength
             is adjusted to the experimental value of $f_\pi$, agrees with the lattice data.
             This behavior can be understood in light of a recent study of corrections beyond RL truncation
             which suggests a near cancellation in the $\rho$-meson of pionic effects and non-resonant corrections from the quark-gluon vertex  \cite{Fischer:2009jm}.
             Setup B provides a description of a quark core which overestimates the experimental values while it approaches
             the lattice results at larger quark masses.

             A comparison to the consistently obtained quark-diquark model result exhibits a discrepancy of only $\sim 5\%$.
             This surprising and reassuring result indicates that a description of the nucleon as a superposition of scalar and axial-vector
             diquark correlations that interact with the remaining quark provides a close approximation to the consistent
             three-quark nucleon amplitude.

    \renewcommand{\arraystretch}{1.0}

             \begin{table}[t]

                              \caption{(adapted from Ref.~\cite{Eichmann:2009qa}) Nucleon masses obtained from the Faddeev equation in setups A and B and
                                       compared to the quark-diquark result. The $\eta$ dependence is indicated for
                                       setup B in parentheses.
                                       }\label{EichmannG_tab:results}

                \begin{center}
                \begin{tabular}{   l @{\;\;} || @{\;\;}l@{\;\;} || @{\;\;}l@{\;\;}   |   @{\;\;} l      }

                                  &  Q-DQ~\cite{Nicmorus:2008vb}      &      Faddeev ($\mathcal{M_A}$)  &  Faddeev ($\mathcal{M_S}$)     \\   \hline

                    Setup A         &  $0.94$                           &     $0.99$                      &   $0.97$                     \\
                    Setup B         &  $1.26(2)$                        &     $1.33(2)$                   &   $1.31(2)$

                \end{tabular}
                \end{center}

        \end{table}

 \renewcommand{\arraystretch}{1.2}

     \section{Conclusions and outlook}\label{EichmannG_sec:FADDEEV:outlook}

             We have provided details on a fully Poincar\'e-covariant three-quark solution of the nucleon's Faddeev equation.
             The nucleon amplitude which is generated by a gluon ladder-exchange is predominantly described by $s$- and $p$-wave Dirac structures,
             and the flavor independence of the kernel leads to a mass degeneracy.
             The resulting nucleon mass is close to the quark-diquark model result which stresses
	     the reliability of previous quark-diquark studies.

             Due to the considerable computational efforts involved, more results and an in-depth investigation with regard to the complete
             set of invariant variables will be presented in the future. %will follow in subsequent publications.
             Further extensions of the present work will include an analogous investigation of
             the $\Delta$-baryon, more sophisticated interaction kernels, e.g.~in view of pionic corrections,
             and ultimately a comprehensive study of baryon resonances.

     \subsection*{Acknowledgements}\label{EichmannG_sec:FADDEEV:acknowledgements}

                We thank C.\,S.~Fischer, M.~Schwinzerl, and R.~Williams for useful discussions.
                This work was supported by the Helmholtz Young Investigator Grant
                VH-NG-332, the Austrian Science Fund FWF under Projects No. P20592-N16,
                P20496-N16, and Doctoral Program No. W1203, and in part by the European Union
		        (HadronPhysics2 project ``Study of strongly interacting matter'').

%%%%%%%%%%%%%%%%%%%%%%%%%%%%%%%%%%%%%%%%%%%%%%%%%%%%%%%%%%%%%%%%%%%%%%%%%%%%%%%%%%%%%%%%%%%%%%%%%%%%%%%%%%%%%%%%%%%%%%%%%%%%%%%%%%%%%%%%%%%%%%%%%%%%%%%%%%%%%%%%%%%%%%%%%%%%%%%
%%%%%%%%%%%%%%%%%%%%%%%%%%%%%%%%%%%%%%%%%%%%%%%%%%%%%%%%%%%%%%%%%%%%%%%%%%%%%%%%%%%%%%%%%%%%%%%%%%%%%%%%%%%%%%%%%%%%%%%%%%%%%%%%%%%%%%%%%%%%%%%%%%%%%%%%%%%%%%%%%%%%%%%%%%%%%%%
%%%%%%%%%%%%%%%%%%%%%%%%%%%%%%%%%%%%%%%%%%%%%%%%%%%%%%%%%%%%%%%%%%%%%%%%%%%%%%%%%%%%%%%%%%%%%%%%%%%%%%%%%%%%%%%%%%%%%%%%%%%%%%%%%%%%%%%%%%%%%%%%%%%%%%%%%%%%%%%%%%%%%%%%%%%%%%%
%%%%%%%%%%%%%%%%%%%%%%%%%%%%%%%%%%%%%%%%%%%%%%%%%%%%%%%%%%%%%%%%%%%%%%%%%%%%%%%%%%%%%%%%%%%%%%%%%%%%%%%%%%%%%%%%%%%%%%%%%%%%%%%%%%%%%%%%%%%%%%%%%%%%%%%%%%%%%%%%%%%%%%%%%%%%%%%

\begin{appendix}

    \section{Numerical implementation} \label{EichmannG_sec:numerics}

               Similar to the analogous case of a two-body Bethe-Salpeter equation, the Faddeev equation \eqref{EichmannG_faddeev:eq} can be viewed as
               an eigenvalue problem for the kernel $\widetilde{K}_{(3)}$:
              \begin{equation}\label{EichmannG_BOUND-STATE:Eigenvalue-Eq}
                  \widetilde{K}_{(3)}(P^2)\,\Psi_i = \lambda_i(P^2) \,\Psi_i\,,
              \end{equation}
              where $P$ is the total momentum of the three-quark bound state and enters the equation as an external parameter.
              Upon projection onto given quantum numbers, the eigenvalues of $\widetilde{K}_{(3)}$  constitute the trajectories $\lambda_i(P^2)$.
              An intersection $\lambda_i(P^2)=1$ at a certain value $P^2=-M_i^2$ reproduces Eq.\,\eqref{EichmannG_faddeev:eq} and therefore corresponds to a potential physical state with mass $M_i$.
              The largest eigenvalue $\lambda_0$ represents the ground state
              of the quantum numbers under consideration and the remaining ones $\lambda_{i\geq 1}$ its excitations;
              the associated eigenvectors $\Psi_i$ are the bound-state amplitudes.
              (Note that in this context one has to keep in mind the possibility of anomalous states in the excitation spectra of bound-state equation solutions~\cite{Ahlig:1998qf}.)
              To obtain the ground-state solution, Eq.~\eqref{EichmannG_BOUND-STATE:Eigenvalue-Eq} is solved via iteration within a 'guess range' $P^2 \in \{ -M_\text{min}^2,\,-M_\text{max}^2 \}$,
              where $M_\text{max}$ is determined from the singularity structure of the quark propagator (see e.g. \cite{Eichmann:2009zx,Oettel:2001kd}).
              Upon convergence of the eigenvalue $\lambda_0(P^2)$ the procedure is repeated for different $P^2$ until $\lambda_0(P^2=-M^2)=1$, thereby defining the nucleon mass $M$.

            From a numerical point of view, it is advantageous to split the Faddeev equation for $\widetilde{K}_{(3)}=KSS$ into an equation for
            a 'wave function' $\Phi=SS\Psi$ which can be evaluated outside the loop integral, and a
            subsequent integration $\Psi = K \Phi$ that is carried out by calling the wave function $\Phi$ with the loop momenta as its arguments:
                \begin{equation*}
                \begin{split}
                    \Phi^{(a)}(p,q,P) &= S(p_b)\,S(p_c)\,\Psi(p,q,P)\,,\\
                    \Psi(p,q,P) &= \frac{1}{\lambda(P^2)} \sum_{a=1}^3 \int_k K^{(a)}(k)\, \Phi^{(a)}(p^{(a)},q^{(a)},P)\,.
                \end{split}
                \end{equation*}
            The index $a=1\dots 3$ denotes the three permutations of the Faddeev kernel, and $\{a,b,c\}$ is an even permutation of $\{1,2,3\}$.
            For a complete Dirac basis the wave function $\Phi^{(a)}$ can be projected onto the same basis elements as the amplitude, cf. Eq.\,\eqref{EichmannG_faddeev:amp},
            and we denote its dressing functions by $\tilde{f}_k^{(a)}$.
           Exploiting the orthogonality relations \eqref{EichmannG_faddeev:orthogonality} yields coupled equations for the amplitude and wave function dressing functions:
                \begin{equation}\label{EichmannG_faddeev:eqcoupled}
                \begin{split}
                    & f_i(s) =  \sum_j \sum_{a=1}^3 \int_k \mathbf{K}^{(a)}_{ij}(s,t)\,\tilde{f}^{(a)}_j\Big(s^{(a)}\Big)\,, \\
                    & \tilde{f}_i^{(a)}(s) = \sum_j \mathbf{G}^{(a)}_{ij}(s)\,f_j(s)\,,
                \end{split}
                \end{equation}
            where the kernel $\mathbf{K}^{(a)}_{ij}$ and quark propagator matrix $\mathbf{G}^{(a)}_{ij}$ are the matrix elements of $K^{(a)}$ and $SS$ upon projection onto
            `outer' basis elements from the left and `inner' basis elements on the right. We abbreviated the five outer momentum variables by $s = \{p^2,q^2,z_0,z_1,z_2\}$
            and the respective inner variables by $s^{(a)}$: they depend on $s$ and
            the four combinations $t:= \{k^2,\,\hat{k}\cdot\hat{P},\,\hat{k}\cdot\hat{p},\,\hat{k}\cdot\hat{q}\}$.

            Eq.~\eqref{EichmannG_faddeev:eqcoupled} involves an iterated (multidimensional) matrix-vector multiplication, where
            in a straightforward implementation $\mathbf{K}$ and $\mathbf{G}$ would be computed in advance
            whereas the $s^{(a)}$ dependence of the $\tilde{f}^{(a)}_j$ must be interpolated in each iteration step.
            The biggest obstacle in such an approach is the memory requirement of the kernel which, for reasonable accuracy, is of the order of Petabytes and
            hence far beyond the capacities of today's computing resources.
            In this respect it is advantageous to split the kernel $\mathbf{K}$ into the following three contributions:
            \begin{equation} \label{EichmannG_kernel-split}
                \mathbf{K}^{(a)}_{ij}(s,t) = \big[\mathbf{K}_1(k)\big]^{\mu\nu}\,\big[\mathbf{K}^{(a)}_2\big]^{\mu\nu}_{ij}\,\big[\mathbf{K}^{(a)}_3(s,t)\big]_{ij}\,,
            \end{equation}
            where $\mathbf{K}_1$ stems from the rainbow-ladder kernel~\eqref{EichmannG_RL:kernel} modulo its Dirac structure and only depends on the gluon momentum $k$;
            $\mathbf{K}_2$ contains all Dirac traces but is momentum-independent; and $\mathbf{K}_3$ carries the remaining loop-momentum dependence inherent in the $\Gamma_i$ of Eq.~\eqref{EichmannG_FADEEV:Gamma_i}
            which enter the basis elements.
            $\mathbf{K}_1$ and $\mathbf{K}_2$ are independent of the baryon momentum $P^2$ and can be calculated and stored outside of the iterated matrix-vector multiplication.
            $\mathbf{K}_3$ is computed in each iteration step anew but includes only simple products of the loop momenta which, in the absence of any Dirac structure, are evaluated comparatively quickly.
            A construction analogous to Eq.~\eqref{EichmannG_kernel-split} can be applied to the propagator matrix $\mathbf{G}^{(a)}_{ij}$ as well;
            however, here the situation is much less severe since no integration and hence no dependence on $t$ is involved.

            Moreover, using the set of orthogonal momenta defined in Eq.~\eqref{EichmannG_orth-momenta2} turns out to be extremely efficient:
            for instance, choosing the momentum alignment
                 \begin{equation}\label{EichmannG_faddeev:orthogonal-unit-vectors}
                 \begin{split}
                     p &=  \sqrt{p^2} \left\{ 0 ,\; 0 ,\; \sqrt{1-z_1^2} ,\; z_1 \right\}, \\
                     q &=  \sqrt{q^2} \left\{ 0 ,\; \sqrt{1-z_2^2}  \sqrt{1-z_0^2} ,\; \sqrt{1-z_2^2} \,z_0 ,\; z_2 \right\}
                 \end{split}
                 \end{equation}
             yields in the baryon's rest frame:
                 \begin{equation*}
                     \widehat{P}     = \left\{  0 ,\, 0 ,\, 0 ,\, 1 \right\}, \quad
                     \widehat{p_T}   = \left\{  0 ,\, 0 ,\, 1 ,\, 0 \right\}, \quad
                     \widehat{q_{t}} = \left\{  0 ,\, 1 ,\, 0 ,\, 0 \right\}.
                 \end{equation*}
            As a consequence, all outer basis elements effectively do not depend on any momentum variable at all,
            and the momentum dependence is solely carried by the inner basis elements which contribute to $\mathbf{K}^{(a)}_{ij}(s,t)$.

            A combination of these strategies greatly reduces the memory demand to $\lesssim 1$ GB.
            The impact on the run time  due to the on-the-fly computation of $\mathbf{K}_3$ %which is now processed on the fly
            is still slightly outweighed by the time consumed to interpolate the dressing functions $\tilde{f}^{(a)}_j$ inside the integral.
            To address this issue, we drop the dependence on the angular variable $z_0 = \widehat{p_T}\cdot\widehat{q_T}$ which,
            from analogy of the quark-diquark model, is expected to be small.
            In addition we perform an expansion into Chebyshev polynomials of the first kind
            in the remaining angles $z_1^{(a)}$, $z_2^{(a)}$ that appear in the wave function coefficients inside the integral.
            The resulting run times are accessible by a parallel cluster.

            The different solutions $\Psi_\mathcal{M_A}$ and $\Psi_\mathcal{M_S}$ of the Faddeev equation are obtained by choosing
            suitable start functions for the amplitudes $f_i(s)$. To arrive at $\Psi_\mathcal{M_A}$, we start from the structure $\mathsf{S}_{11}^+$
            and set all other amplitudes to zero; for $\Psi_\mathcal{M_S}$ we use the initial amplitude $\mathsf{A}_{11}^+$.
            The full structure of the nucleon amplitude in either case is subsequently generated by iterating Eq.~\eqref{EichmannG_faddeev:eqcoupled}.

%              The current-quark mass is an input to the quark DSE \eqref{a}
%              and can be mapped onto the pion mass upon solving the pseudoscalar meson BSE.
%              This allows for a determination of all subsequent results as a function of $m_\pi^2$, where the physical point is characterized by $m_\pi=138$ MeV.
%              Varying the current mass, and thus the pion mass, enables a direct comparison to lattice data and their chiral extrapolations.
%
%

     \section{Angular momentum decomposition} \label{EichmannG_sec:partialwave}

             The Dirac basis elements $\mathsf{X}_{ij,k}^r$ in Table~\ref{EichmannG_tab:faddeev:basis} can be classified with respect to their quark-spin and orbital angular momentum content in the nucleon's rest frame.
             Only the total angular momentum $j=1/2$ of the nucleon is Poincar\'e-invariant  while the interpretation in terms of total quark spin
             and orbital angular momentum will differ in every frame.
             The spin is described by the Pauli-Lubanski operator:
             \begin{equation}
                 W^\mu = \frac{1}{2} \,\epsilon^{\mu\nu\alpha\beta} \hat{P}^\nu J^{\alpha\beta}\,,
             \end{equation}
             where we chose the total momentum $P$ to be normalized.
             $J^{\mu\nu}$ and $P^\mu$ are the generators of the Poincar\'e algebra satisfying the usual commutation relations.
             The eigenvalues of the square of the Pauli-Lubanski operator,
             \begin{equation}
                 W^2 = \frac{1}{2} J^{\mu\nu}J^{\mu\nu} + \hat{P}^\mu \hat{P}^\nu J^{\mu\alpha} J^{\alpha\nu}  \longrightarrow j(j+1)
             \end{equation}
             define the spin $j$ of the particle.
%             is one of the two Casimir operators of the Poincar\'e group. % which describe the mass and spin of a particle.
%             Its eigenvalues are given by $W^2 \longrightarrow j(j+1)$, where $j$ describes the spin of the particle.
%
%
             For a system of three particles with total momentum $P$ and relative momenta $p$ and $q$,
             the total angular momentum operator consists of the total quark spin $\vect{S}$ and the relative orbital angular momentum $\vect{L}=\vect{L}_{(p)}+\vect{L}_{(q)}$.
             Subsuming them into Lorentz-covariant operators
             \begin{align}
                 S^{\mu} &= \textstyle\frac{1}{4} \displaystyle \epsilon^{\mu\nu\alpha\beta} \hat{P}^\nu
                            \left( \sigma^{\alpha\beta} \otimes \mathds{1} \otimes \mathds{1} +  \text{perm.} \right), \nonumber \\
%                            \mathds{1} \otimes \sigma^{\alpha\beta} \otimes \mathds{1} + \mathds{1} \otimes \mathds{1} \otimes \sigma^{\alpha\beta} \right), \\
                 L_{(p)}^{\mu} &= \textstyle\frac{i}{2} \displaystyle \epsilon^{\mu\nu\alpha\beta} \hat{P}^\nu \left( p^\alpha \partial_p^\beta - p^\beta \partial_p^\alpha \right) \mathds{1} \otimes \mathds{1} \otimes \mathds{1}, \\
                 L_{(q)}^{\mu} &= \textstyle\frac{i}{2} \displaystyle \epsilon^{\mu\nu\alpha\beta} \hat{P}^\nu \left( q^\alpha \partial_q^\beta - q^\beta \partial_q^\alpha \right) \mathds{1} \otimes \mathds{1} \otimes \mathds{1}  \nonumber
             \end{align}
             with $W^\mu = S^\mu + L_{(p)}^\mu + L_{(q)}^\mu$ yields the identities
             \begin{align*}
                 S^2 &= \textstyle\frac{9}{4} \displaystyle \,\mathds{1} \otimes \mathds{1} \otimes \mathds{1} + \textstyle\frac{1}{4} \displaystyle \left( \sigma_T^{\mu\nu} \otimes \sigma_T^{\mu\nu} \otimes \mathds{1} + \text{perm.} \right)\,, \\[0.1cm]
                 S \!\cdot\! L_{(p)} &= \textstyle\frac{i}{2} \displaystyle \,\widetilde{p}^\mu\, \widetilde{\partial}_p^\nu \left( \sigma_T^{\mu\nu} \otimes  \mathds{1} \otimes \mathds{1} + \text{perm.} \right)\,, \\[0.15cm]
                 L_{(p)}^2 &= 2  \, \widetilde{p}\cdot\widetilde{\partial}_p + \widetilde{p}^\mu \,\widetilde{p}^\nu \,\widetilde{\partial}_p^\mu \widetilde{\partial}_p^\nu - \widetilde{p}^2  \,\widetilde{\partial}_p^2\,,\\[0.1cm]
                 L_{(p)}\!\cdot\! L_{(q)} &=  \widetilde{p}^\mu\,\widetilde{q}^\nu\, \widetilde{\partial}_p^\nu \,\widetilde{\partial}_q^\mu -(\widetilde{p}  \cdot  \widetilde{q}) \,(\widetilde{\partial}_p  \cdot \widetilde{\partial}_q)\,,
             \end{align*}
             where we abbreviated
             \begin{equation}
             \begin{array}{l} \widetilde{p}^\mu = T^{\mu\nu}_P\,p^\nu\,, \\ \widetilde{q}^\mu = T^{\mu\nu}_P\,q^\nu\,, \end{array} \quad
             \begin{array}{l} \widetilde{\partial}_p^\mu = T^{\mu\nu}_P\,\partial/\partial p^\nu\,, \\ \widetilde{\partial}_q^\mu = T^{\mu\nu}_P\,\partial/\partial q^\nu\,, \end{array} \quad
             \end{equation}
             and $\sigma_T^{\mu\nu} = -i\,[\gamma_T^\mu,\gamma_T^\nu]/2$.
             From these relations one can verify that the $\mathsf{X}_{ij,k}^r$ are
             eigenfunctions of $W^2$ with $j=1/2$
             as well as eigenfunctions of $S^2 \longrightarrow s(s+1)$ and $L^2\longrightarrow l(l+1)$
             which, in the nucleon's rest frame (where $L^2 = \vect{L}^2$ and $S^2 = \vect{S}^2$), assume the interpretation of total quark spin and intrinsic quark orbital angular momentum.
             Such a partial-wave decomposition yields 32 basis elements for each total quark spin $s=1/2$ and $s=3/2$, respectively,
             and a classification into sets of 8 $s$ waves ($l=0$), 36 $p$ waves ($l=1$), and 20 $d$ waves ($l=2$).

          \renewcommand{\arraystretch}{1.2}

              \section{Multispinor notation} \label{EichmannG_sec:threespinor}

             The basis elements $\mathsf{X}_{ij,k}^r$ can be expressed in terms of %quark three-spinors frequently used in the literature, e.g. Ref.\,\cite{Carimalo:1992ia}.
              the Dirac spinors $U^\sigma(P)$, $V^\sigma(P) = \gamma_5 U^\sigma(P)$ which satisfy the free Dirac equation for a spin-$\nicefrac{1}{2}$ particle, i.e.
              which are eigenspinors of the positive and negative energy projectors $\Lambda_\pm$.
              Normalized to $\conjg{U}^\rho(P) \,U^\sigma(P) = \delta_{\rho\sigma}$, they are given by
              \begin{equation}\label{EichmannG_NUCLEON:DiracSpinors}
                  U^\sigma(P) = \sqrt{\frac{\varepsilon+M}{2\,M}} \left( \begin{array}{c} w^\sigma \\ \textstyle\frac{\vect{\scriptstyle\sigma}\cdot \vect{\scriptstyle P}}{\varepsilon+M}\,w^\sigma \!\!\end{array}\right), \quad
                  \begin{array}{rcl} w^\uparrow &=& (1,0) \\ w^\downarrow &=& (0,1) \end{array}
              \end{equation}
              with $\varepsilon=\sqrt{\vect{P}^2+M^2}$.
              A 64-dimensional basis for the nucleon wave function was constructed in Ref.~\cite{Carimalo:1992ia} from the quark three-spinors
              $UUU$, $VVU$, $VUV$, $UVV$ and their parity-reversed counterparts $VVV$, $UUV$, $UVU$, and $VUU$
              by equipping them with the 8 possible spin-up/down arrangements $\uparrow\uparrow\uparrow$, $\uparrow\uparrow\downarrow$, and so on.
              Identities such as
              \begin{equation}
                  \Lambda_+ = U^\uparrow \conjg{U}^\uparrow + U^\downarrow \conjg{U}^\downarrow\,, \quad  \Lambda_+ \gamma_5 C = U^\downarrow U^\uparrow - U^\uparrow U^\downarrow
              \end{equation}
              allow to derive the following relations for the momentum-independent $s$-wave basis states of Table~\ref{EichmannG_tab:faddeev:basis}:
              \begin{equation*}\label{EichmannG_faddeev:spinor1}
              \begin{array}{rl}
                 -\mathsf{S}_{11}^+ \, U^\uparrow &=  (U^\uparrow  U^\downarrow - U^\downarrow  U^\uparrow) \, U^\uparrow\,,  \\
                 -\mathsf{S}_{11}^- \, U^\uparrow &=  (V^\uparrow  V^\downarrow - V^\downarrow  V^\uparrow) \, U^\uparrow\,,  \\
                 -\mathsf{P}_{11}^+ \, U^\uparrow &=  (V^\uparrow  U^\downarrow - V^\downarrow  U^\uparrow) \, V^\uparrow\,,  \\
                 -\mathsf{P}_{11}^- \, U^\uparrow &=  (U^\uparrow  V^\downarrow - U^\downarrow  V^\uparrow) \, V^\uparrow
              \end{array}
              \end{equation*}
              and
              \begin{equation*}\label{EichmannG_faddeev:spinor2}
              \begin{array}{rl}
                  \mathsf{A}_{11}^+ \, U^\uparrow &=  (U^\uparrow  U^\downarrow + U^\downarrow  U^\uparrow) \, U^\uparrow - 2 \, U^\uparrow  U^\uparrow  U^\downarrow\,,  \\
                 -\mathsf{A}_{11}^- \, U^\uparrow &=  (V^\uparrow  V^\downarrow + V^\downarrow  V^\uparrow) \, U^\uparrow - 2 \, V^\uparrow  V^\uparrow  U^\downarrow\,,   \\
                  \mathsf{V}_{11}^+ \, U^\uparrow &=  (V^\uparrow  U^\downarrow + V^\downarrow  U^\uparrow) \, V^\uparrow - 2 \, V^\uparrow  U^\uparrow  V^\downarrow\,,   \\
                 -\mathsf{V}_{11}^- \, U^\uparrow &=  (U^\uparrow  V^\downarrow + U^\downarrow  V^\uparrow) \, V^\uparrow - 2 \, U^\uparrow  V^\uparrow  V^\downarrow\,.
              \end{array}
              \end{equation*}
              Due to the Poincar\'e-covariant construction of Eqs.\,(\ref{EichmannG_basisSP}-\ref{EichmannG_basisX}), the remaining 56 covariants depend on the relative momenta.
              For instance, the Dirac structure $\mathsf{S}_{13}^-$ satisfies the relation
              \begin{equation}\label{EichmannG_FE:S13}
                  i\,\mathsf{S}_{13}^- \, U^\uparrow =  (V^\uparrow  V^\downarrow - V^\downarrow  V^\uparrow) \, \left( \left([\widehat{p_T}]_1 + i [\widehat{p_T}]_2 \right) V^\downarrow + [\widehat{p_T}]_3 V^\uparrow  \right)\,.
              \end{equation}
              Using the special momentum alignment \eqref{EichmannG_faddeev:orthogonal-unit-vectors} leads to
              \begin{equation}\label{EichmannG_FADDEEV:S13}
                  i\,\mathsf{S}_{13}^- \, U^\uparrow \stackrel{\eqref{EichmannG_faddeev:orthogonal-unit-vectors}}{\longlongrightarrow}  (V^\uparrow  V^\downarrow - V^\downarrow  V^\uparrow) \, V^\uparrow \,
              \end{equation}
              and similar relations for the remaining basis elements in Table~\ref{EichmannG_tab:faddeev:basis}.
              From a three-spinor perspective, Eq.~\eqref{EichmannG_FADDEEV:S13} is at the same time the 'parity-flipped' counterpart of $\mathsf{S}_{11}^+$,
              \begin{equation}
                  -(\gamma_5 \otimes \gamma_5) \,\mathsf{S}_{11}^+  \,(\gamma_5^T \otimes U^\uparrow) = (V^\uparrow  V^\downarrow - V^\downarrow  V^\uparrow) \, V^\uparrow \,,
              \end{equation}
              which would appear in the amplitude of a negative-parity nucleon ($1/2^-$) and that of a positive-parity antinucleon ($\overline{1/2^+}$).
              Nevertheless this seemingly odd-parity structure still contributes to the ($1/2^+$) state through Eq.\,\eqref{EichmannG_FE:S13},
              where the parity flip induced by the spinor replacement
              $U \rightarrow V$ is saturated by an odd parity carried by the relative momentum $\widehat{p_T}$.
              As a consequence, indeed all 64 three-spinor combinations contribute to the nucleon's amplitude.

\end{appendix}

\bibliographystyle{epj}
%\bibliography{lit}
\bibliography{lit}

\begin{thebibliography}{48}

\bibitem{Faddeev:1960su}
L.D. Faddeev, Sov. Phys. JETP \textbf{12}, 1014 (1961)
%%CITATION = SPHJA,12,1014;%%

\bibitem{Gloeckle:1995jg}
W.~Gloeckle, H.~Witala, D.~Huber, H.~Kamada, J.~Golak, Phys. Rept.
  \textbf{274}, 107 (1996)
%%CITATION = PRPLC,274,107;%%

\bibitem{Salpeter:1951sz}
E.E. Salpeter, H.A. Bethe, Phys. Rev. \textbf{84}, 1232 (1951)
%%CITATION = PHRVA,84,1232;%%

\bibitem{Taylor:1966zza}
J.G. Taylor, Phys. Rev. \textbf{150}, 1321 (1966)
%%CITATION = PHRVA,150,1321;%%

\bibitem{Boehm:1976ya}
M.~Boehm, R.F. Meyer, Annals Phys. \textbf{120}, 360 (1979)
%%CITATION = APNYA,120,360;%%

\bibitem{Loring:2001kv}
U.~Loring, K.~Kretzschmar, B.C. Metsch, H.R. Petry, Eur. Phys. J. \textbf{A10},
  309 (2001)
%%CITATION = HEP-PH/0103287;%%

\bibitem{Fischer:2006ub}
C.S. Fischer, J. Phys. \textbf{G32}, R253 (2006)
%%CITATION = HEP-PH/0605173;%%

\bibitem{Roberts:2007jh}
C.D. Roberts, M.S. Bhagwat, A.~Holl, S.V. Wright, Eur. Phys. J. ST
  \textbf{140}, 53 (2007)
%%CITATION = 0802.0217;%%

\bibitem{Machida:1974xw}
S.~Machida, H.~Nakkagawa, Prog. Theor. Phys. \textbf{54}, 243 (1975)
%%CITATION = PTPKA,54,243;%%

\bibitem{Henriques:1975uh}
A.B. Henriques, B.H. Kellett, R.G. Moorhouse, Ann. Phys. \textbf{93}, 125
  (1975)
%%CITATION = APNYA,93,125;%%

\bibitem{Weber:1986qw}
H.J. Weber, Ann. Phys. \textbf{177}, 38 (1987)
%%CITATION = APNYA,177,38;%%

\bibitem{Beyer:1998xy}
M.~Beyer, C.~Kuhrts, H.J. Weber, Annals Phys. \textbf{269}, 129 (1998)
%%CITATION = NUCL-TH/9804021;%%

\bibitem{Karmanov:1998jp}
V.A. Karmanov, Nucl. Phys. \textbf{A644}, 165 (1998)
%%CITATION = NUCL-TH/9802053;%%

\bibitem{Sun:2001ir}
X.P. Sun, H.J. Weber, Int. J. Mod. Phys. \textbf{A17}, 2535 (2002)
%%CITATION = HEP-PH/0102240;%%

\bibitem{Carimalo:1992ia}
C.~Carimalo, J. Math. Phys. \textbf{34}, 4930 (1993)

\bibitem{Kielanowski:1979eb}
P.~Kielanowski, Z. Phys. \textbf{C3}, 267 (1979)
%%CITATION = ZEPYA,C3,267;%%

\bibitem{Falkensteiner:1981ab}
P.~Falkensteiner, Zeit. Phys. \textbf{C11}, 343 (1982)
%%CITATION = ZEPYA,C11,343;%%

\bibitem{Stadler:1997iu}
A.~Stadler, F.~Gross, M.~Frank, Phys. Rev. \textbf{C56}, 2396 (1997)
%%CITATION = NUCL-TH/9703043;%%

\bibitem{Wick:1954eu}
G.C. Wick, Phys. Rev. \textbf{96}, 1124 (1954)
%%CITATION = PHRVA,96,1124;%%

\bibitem{Cutkosky:1954ru}
R.E. Cutkosky, Phys. Rev. \textbf{96}, 1135 (1954)
%%CITATION = PHRVA,96,1135;%%

\bibitem{Karmanov:2008bx}
V.A. Karmanov, P.~Maris, Few Body Syst. \textbf{46}, 95 (2009)
%%CITATION = 0811.1100;%%

\bibitem{Anselmino:1992vg}
M.~Anselmino, E.~Predazzi, S.~Ekelin, S.~Fredriksson, D.B. Lichtenberg, Rev.
  Mod. Phys. \textbf{65}, 1199 (1993)
%%CITATION = RMPHA,65,1199;%%

\bibitem{Hellstern:1997pg}
G.~Hellstern, R.~Alkofer, M.~Oettel, H.~Reinhardt, Nucl. Phys. \textbf{A627},
  679 (1997)
%%CITATION = HEP-PH/9705267;%%

\bibitem{Oettel:2000jj}
M.~Oettel, R.~Alkofer, L.~von Smekal, Eur. Phys. J. \textbf{A8}, 553 (2000)
%%CITATION = NUCL-TH/0006082;%%

\bibitem{Alkofer:2004yf}
R.~Alkofer, A.~Holl, M.~Kloker, A.~Krassnigg, C.D. Roberts, Few Body Syst.
  \textbf{37}, 1 (2005)
%%CITATION = NUCL-TH/0412046;%%

\bibitem{Holl:2005zi}
A.~Holl, R.~Alkofer, M.~Kloker, A.~Krassnigg, C.D. Roberts, S.V. Wright, Nucl.
  Phys. \textbf{A755}, 298 (2005)
%%CITATION = NUCL-TH/0501033;%%

\bibitem{Eichmann:2007nn}
G.~Eichmann, A.~Krassnigg, M.~Schwinzerl, R.~Alkofer, Annals Phys.
  \textbf{323}, 2505 (2008)
%%CITATION = 0712.2666;%%

\bibitem{Eichmann:2008ef}
G.~Eichmann, I.C. Cloet, R.~Alkofer, A.~Krassnigg, C.D. Roberts, Phys. Rev.
  \textbf{C79}, 012202 (2009)
%%CITATION = 0810.1222;%%

\bibitem{Nicmorus:2008vb}
D.~Nicmorus, G.~Eichmann, A.~Krassnigg, R.~Alkofer, Phys. Rev. \textbf{D80},
  054028 (2009)
%%CITATION = 0812.1665;%%

\bibitem{Eichmann:2008kk}
G.~Eichmann, R.~Alkofer, A.~Krassnigg, D.~Nicmorus, PoS \textbf{CONFINEMENT8},
  077 (2008)
%%CITATION = 0812.3183;%%

\bibitem{Nicmorus:2008eh}
D.~Nicmorus, G.~Eichmann, A.~Krassnigg, R.~Alkofer, PoS \textbf{CONFINEMENT8},
  052 (2008)
%%CITATION = 0812.2966;%%

\bibitem{Maris:2002yu}
P.~Maris, Few Body Syst. \textbf{32}, 41 (2002)
%%CITATION = NUCL-TH/0204020;%%

\bibitem{Maris:2004bp}
P.~Maris, Few Body Syst. \textbf{35}, 117 (2004)
%%CITATION = NUCL-TH/0409008;%%

\bibitem{Alkofer:2005ug}
R.~Alkofer, M.~Kloker, A.~Krassnigg, R.F. Wagenbrunn, Phys. Rev. Lett.
  \textbf{96}, 022001 (2006)
%%CITATION = HEP-PH/0510028;%%

\bibitem{Eichmann:2009qa}
G.~Eichmann, R.~Alkofer, A.~Krassnigg, D.~Nicmorus (2009), arXiv:0912.2246
%%CITATION = 0912.2246;%%

\bibitem{Eichmann:2009zx}
G.~Eichmann, Ph.D. thesis, University of Graz  (2009), arXiv:0909.0703
%%CITATION = 0909.0703;%%

\bibitem{Maris:1997hd}
P.~Maris, C.D. Roberts, P.C. Tandy, Phys. Lett. \textbf{B420}, 267 (1998)
%%CITATION = NUCL-TH/9707003;%%

\bibitem{Holl:2004fr}
A.~Holl, A.~Krassnigg, C.D. Roberts, Phys. Rev. \textbf{C70}, 042203 (2004)
%%CITATION = NUCL-TH/0406030;%%

\bibitem{Krassnigg:2009zh}
A.~Krassnigg, Phys. Rev. \textbf{D80}, 114010 (2009)
%%CITATION = 0909.4016;%%

\bibitem{Maris:1999nt}
P.~Maris, P.C. Tandy, Phys. Rev. \textbf{C60}, 055214 (1999)
%%CITATION = NUCL-TH/9905056;%%

\bibitem{Eichmann:2008ae}
G.~Eichmann, R.~Alkofer, I.C. Cloet, A.~Krassnigg, C.D. Roberts, Phys. Rev.
  \textbf{C77}, 042202 (2008)
%%CITATION = 0802.1948;%%

\bibitem{Fischer:2009jm}
C.S. Fischer, R.~Williams, Phys. Rev. Lett. \textbf{103}, 122001 (2009)
%%CITATION = 0905.2291;%%

\bibitem{AliKhan:2003cu}
A.~Ali~Khan et~al. (QCDSF-UKQCD), Nucl. Phys. \textbf{B689}, 175 (2004)
%%CITATION = HEP-LAT/0312030;%%

\bibitem{Frigori:2007wa}
R.~Frigori et~al., PoS \textbf{LAT2007}, 114 (2007)
%%CITATION = ARXIV:0709.4582;%%

\bibitem{AliKhan:2001tx}
A.~Ali~Khan et~al. (CP-PACS), Phys. Rev. \textbf{D65}, 054505 (2002)
%%CITATION = HEP-LAT/0105015;%%

\bibitem{Allton:2005fb}
C.R. Allton, W.~Armour, D.B. Leinweber, A.W. Thomas, R.D. Young, Phys. Lett.
  \textbf{B628}, 125 (2005)
%%CITATION = HEP-LAT/0504022;%%

\bibitem{Ahlig:1998qf}
S.~Ahlig, R.~Alkofer, Annals Phys. \textbf{275}, 113 (1999)
%%CITATION = HEP-TH/9810241;%%

\bibitem{Oettel:2001kd}
M.~Oettel, L.~Von~Smekal, R.~Alkofer, Comput. Phys. Commun. \textbf{144}, 63
  (2002)
%%CITATION = HEP-PH/0109285;%%

\end{thebibliography}

%\begin{thebibliography}{99}
%\bibitem{SchmidtPL_RefJ}
%A.A.~Author, Journal \textbf{Volume}, (year) page numbers
%\bibitem{SchmidtPL_RefB}
%B.B.~Author, \textit{Book title} (Publisher, place year) page numbers
%\bibitem{SchmidtPL_RefNP}
%N. Surname1 and N. Surname2, Nucl. Phys. \textbf{Xnum},
%(year) page numbers
%\bibitem{SchmidtPL_RefPR}
%N.N. Surname3 \textit{et al.}, Phys. Rev. \textbf{Xnum} (year) page
%\bibitem{SchmidtPL_RefPL}
%N. Surname4, N.N. Surname5, and N. Surname6, Phys. Lett.
%\textbf{Xnum} (year) page numbers
%\bibitem{SchmidtPL_RefProc}
%N.N. Surname7, \textit{Proc. of the TitleOfProceedings} (eds. NameOfEditors,
%place, year) page numbers.
%\end{thebibliography}

\end{document}